\documentclass[twocolumn]{aastex63}

\usepackage{newtxtext,newtxmath}
\usepackage[T1]{fontenc}  \usepackage{ae,aecompl}
\usepackage{graphicx} \usepackage{amsmath}     \usepackage{amssymb}
\usepackage{epsfig}  \usepackage{lineno}  \usepackage{overpic}
\usepackage{hyperref}
 
\newcommand{\msolar}{M$_\odot$} \newcommand{\Lsolar}{L$_\odot$}

 \newcommand{\COtw}{$^{12}\rm CO$} \newcommand{\COth}{$^{13}\rm CO$} \newcommand{\CeiO}{$\rm C^{18}O$}

  \newcommand{\Lwc}{L$_{12\micron}$}

\received{}  \revised{}  \accepted{} \submitjournal{ApJ}

\shorttitle{\textit{WISE} 12 $\mu$m  vs. molecular gas in the MALATANG sample}
\shortauthors{Y. Gao et al.}

\begin{document}

\title{The correlation between \textit{WISE} 12 \micron\ emission and molecular gas tracers on sub-kpc scales in nearby star-forming galaxies}

\correspondingauthor{Yang Gao}  \email{gaoyang@pmo.ac.cn}

\author{Yang Gao} \affiliation{Purple Mountain Observatory \&
  Key Lab. of Radio Astronomy, Chinese Academy of Sciences, 10 Yuanhua Road, Nanjing
  210023, China}

\author{Qing-Hua Tan} \affiliation{Purple Mountain Observatory \& Key
  Lab. of Radio Astronomy, Chinese Academy of Sciences, 10 Yuanhua Road, Nanjing
  210023, China}
  
\author{Yu Gao} \affiliation{Department of Astronomy, Xiamen University, Xiamen, Fujian 361005, China}
\affiliation{Purple Mountain Observatory \& Key
  Lab. of Radio Astronomy, Chinese Academy of Sciences, 10 Yuanhua Road, Nanjing
  210023, China}

\author{Min Fang} \affiliation{Purple Mountain Observatory \& Key
  Lab. of Radio Astronomy, Chinese Academy of Sciences, 10 Yuanhua Road, Nanjing
  210023, China}
  
\author{Ryan Chown} \affiliation{Department of Physics \& Astronomy, University of Western Ontario, London, ON N6A 3K7, Canada}

\author{Qian Jiao} \affiliation{School of Electrical and Electronic Engineering, Wuhan Polytechnic University, Wuhan 430023, China}
\affiliation{Purple Mountain Observatory \& Key
  Lab. of Radio Astronomy, Chinese Academy of Sciences, 10 Yuanhua Road, Nanjing
  210023, China}
  
\author{Chun-Sheng Luo} \affiliation{Purple Mountain Observatory \& Key
  Lab. of Radio Astronomy, Chinese Academy of Sciences, 10 Yuanhua Road, Nanjing
  210023, China}

\begin{abstract}
 
We complement the MALATANG sample of dense gas in nearby galaxies with archival observations of \COtw\ and its isotopologues to determine scaling relations between \textit{Wide-field Infrared Survey Explorer} (\textit{WISE}) 12 {\micron} emission and molecular gas tracers at sub-kiloparsec scales. We find that 12 {\micron}  luminosity is more tightly correlated with \COtw\ than it is with \COth\ or dense gas tracers. Residuals between predicted and observed \COtw\ are only weakly correlated with molecular gas mass surface density ($\Sigma_{\rm mol}$) in regions where $\Sigma_{\rm mol}$ is very low ($\sim 10~{\rm M_{\odot}~pc^{-2}}$).  Above this limit, the \COtw\ residuals show no correlations with physical conditions of molecular gas, while \COth\ residuals depend on the gas optical depth and temperature. 
By analyzing differences from galaxy to galaxy, we confirm that the \COtw-12 {\micron} relation is strong and statistically robust with respect to star forming galaxies and AGN hosts. These results suggest that \textit{WISE} 12 {\micron} emission can be used to trace total molecular gas instead of dense molecular gas, likely because polycyclic aromatic hydrocarbons (PAHs, a major contributor to \textit{WISE} 12 \micron~emission) may be well-mixed with the gas that is traced by \COtw.
We propose that \textit{WISE} 12 {\micron} luminosity can be used to estimate molecular gas surface density for statistical analyses of the star formation process in galaxies.

\end{abstract} 



\section{Introduction}
\label{sec:introduction}

In current galaxy formation models, galaxies form within 
dark matter halos. Some of the gas in the potential wells of these halos is able to cool, condense, gravitationally collapse, and then form stars \citep[e.g.][]{White1978}.  It is thus crucial to understand the
physical processes which govern how gas is converted into stars in galaxies before one can have a complete picture of galaxy formation and evolution. Since the first power-law relation studying these process provided by \citet{Schmidt1959}, the most commonly used Kennicutt–Schmidt (KS) law relates the surface densities of star formation rate (SFR) and that of cold gas in a galactic disc, which is characterized as:
$\Sigma_\mathrm{SFR} \propto \Sigma(\mathrm{gas})^{1.4}$ \citep{Kennicutt1998}, where gas includes both atomic and molecular gas. Thanks to observations of multiple molecular species at (sub)millimeter bands, 
it is found that star formation takes place in the clumps and cores within giant molecular clouds (GMCs), so SFR is linearly correlated with the amount of molecular gas (H$_{2}$, traced by CO) instead of the atomic gas \citet{Baan2008,Bigiel2008,Leroy2008,Kennicutt2012}. 

However, the physical processes that produce the empirical correlation between SFR and gas
are far from being fully understood. For instance, the $\Sigma_\mathrm{SFR}$-$\Sigma_\mathrm{H_{2}}$ law may not be universal from both observations \citep{Daddi2010,Shetty2013} and theoretical predictions \citep{Krumholz2005,Elmegreen2015,Elmegreen2018}, and the existing stars may play an  important role in regulating SFR in regions or galaxies with low gas surface densities \citep{Shi2011,Shi2018}. Meanwhile, \citet{Gao2004a,Gao2004b} revealed a tight linear relation between the integrated SFRs (traced by total infrared luminosity) and dense molecular gas masses (derived from HCN emission) of normal and starburst galaxies. This relation still hold at the scale of Galactic massive cores in the Milky Way \citep{Wu2005}, where the KS law breaks down \citep{Onodera2010,Nguyen-Luong2016}.

From an observational perspective, large optical imaging and spectroscopic surveys have in the past two decades well established that the cessation of star formation is one of the driving processes of galaxy evolution over the past 8–10 Gyr \citep{Bell2004,Bundy2006,Faber2007}. However, despite a rich history of studies, a full understanding of the way in which star formation shuts down remains elusive. No matter what drives the star formation cessation in galaxies, the cold gas supply must be cut off or (if the galaxies manage to retain their cold gas) the star formation efficiency must be effectively reduced. In any case, it is critical to understand the cold gas content (especially molecular gas) and its relationship to star formation and galaxy properties.

At present, samples of galaxies with molecular gas measurements are far smaller and have poorer spatial resolution compared to optical surveys, which limits the sample sizes of studies of molecular gas and star formation in nearby galaxies.
\citet{Jiang2015,Gao2019,Chown2021} claim that the strong and tight relation between CO luminosity and \textit{WISE} 12 
\micron\ (hereafter W3)  luminosity on both galaxy and sub-galaxy scales should be applied to large samples of galaxies to enable star formation and gas-related processes to be studied statistically. At the moment, one advantage that \textit{WISE} has over the \textit{James Webb Space Telescope} (\textit{JWST}) is a full-sky survey, which makes \textit{WISE}-based molecular gas estimators viable over the entire sky.
\citet{Leroy2021} also confirm a strong, linear correlation between W3 emission and CO(2-1) intensity, 
and adopt \textit{WISE} 12 \micron\ emission as the best template to derived the aperture correction for CO(2-1) luminosity in PHANGS–ALMA observations. 
However, the physical origin (if any) is still under question, though the relation is very tight.

The \textit{WISE} 12 \micron\ band spanning a broad wavelength range from 7.5 to 16.5 \micron\ \citep{Jarrett2011}, includes prominent polycyclic aromatic hydrocarbon (PAH) emission, and other dust emission \citep{Wright2010}.  The PAH features at 8.6, 11.3 and 12.7 \micron\ are produced by C--H bending modes in and out of plane, and 7.7 \micron\ PAH emission (which partly contributes to W3 emission) is from C–C stretching modes \citep{Draine2007}.
PAH emission and warm dust emission are associated with star formation \citep{Xie2019,Popescu2000}, so 12 \micron\ emission could be a reliable SFR indicator as  shown by \citet{Cluver2017}, and then is linked to the raw material of star
formation, molecular gas (thus CO emission). 

Meanwhile an emerging and direct PAH–CO luminosity relation has been reported \citep{Cortzen2019}. It is still debated where and how PAHs form, however, both observational and theoretical studies suggest that, closer to the star, PAHs can be effectively destroyed, fragmented, or ionized by intense and hard UV photons with reduced shielding by dust \citep{Boulanger1988,Giard1994,Allain1996,Povich2007,Sandstrom2012}. 
Bright PAH emission is only visible from shell-like structures on the surfaces of H\;\textsc{ii} regions \citep{Churchwell2006,Rho2006,Sandstrom2012,Schinnerer2013}.
So PAH carriers preferentially reside in the regions dominated by cold dust and molecular clouds, where deservedly they should share similar excitation mechanisms (mainly by the interstellar radiation field) \citep{Haas2002,Bendo2008,Bendo2010,Sandstrom2010,Sandstrom2012}.
This conjecture is in agreement with the model \citep{Akimkin2015}, where PAHs with a non-zero charge are well coupled to the gas due to the largest surface-to-mass ratio. 
Besides, CO and dust continuum always trace the same gas component, even in $z=6.6$ quasar host galaxy \citep{Li2022}.
So there's a reasonable prospect that 12 {\micron} emission is a better predictor of molecular gas than SFR.

It will be helpful taking into account the multi-phase (dense) nature of molecular gas to understand the physical connections between molecular gas, 12 \micron\ emission and SFR.
MALATANG (Mapping the dense molecular gas in the strongest star-forming galaxies), a large program on the James Clerk Maxwell Telescope (JCMT), is the first systematic survey to provide deep HCN(4-3) and HCO$^{+}$(4-3) maps (tracing the densest gas undergoing star formation) in the largest sample of nearby galaxies \citep{Tan2018}.
The first phase of MALATANG (project code: M16AL007) survey observed 23 nearby star-forming galaxies beyond the local group, which were selected from the \textit{Infrared Astronomical Satellite} (\textit{IRAS}) Revised Bright Galaxy Sample \citep{Sanders2003} with $f_{\mbox{60\micron}}$\textgreater\ 50 Jy and $f_{\mbox{100\micron}}$\textgreater\ 100 Jy. To improve the sensitivity and serve as a synergy study with The “EMIR Multiline Probe of the ISM Regulating Galaxy Evolution” \citep[EMPIRE;][]{Bigiel2016,Jimenez-Donaire2019}, which used the IRAM 30-m telescope to map multiple molecular lines of nine nearby, face-on massive spiral galaxies, the second phase (project code: M20AL022) selected a sample of 11 galaxies, with six already detected in the first phase and the other five from EMPIRE sample that were not observed in the first phase. In total, there are 28 IR-bright star-forming galaxies in MALATANG sample. The second phase of MALATANG has started the observations since April 2021 and is still underway.
The observations of abundant data molecular lines in these very nearby galaxies makes
MALATANG ideal laboratory for studying the relation between molecular gas and 12 \micron\ emission
down to very small scales of $\sim$ 1 kpc or less.
There are 8 AGNs in the sample, which allow us to study the effect of AGNs on the spatially resolved relation.

We complement MALATANG sample with existing observations of \COtw\ and its isotopologues from the literature in order to characterize scaling relations between gas tracer and 12 \micron\ luminosity.
The purpose of our work is threefold. First, we revisit the $L_{\rm gas}$-$L_{\mbox{12\micron}}$ relation on sub-kpc scales. Second, we explore the physical explanation behind them by comparing the relation of various molecular gas tracers and analyzing offsets from the main relations. Third, we test the applicability of molecular gas mass estimators based on  12 \micron\ luminosity in spatially resolved regions and different galaxies.

Our paper is organized as follows. In the next section we describe the data used in this paper. In Section~\ref{sec:gas_w3} we show the correlations between luminosities of various molecular gas tracers and W3 luminosities, and examine the dependence of offset (scatter) around on physical conditions of molecular gas. In Section~\ref{sec:discussion}, we quantify the variation in the relations between different galaxy, and discuss the robustness of estimated \COtw.  Finally, we summarize our work in Section~\ref{sec:summary}.

\section{Data}

\label{sec:data}
\subsection{CO Data}
\label{sec:co}

The CO data are mainly from two large programs: 
CO Multi-line Imaging of Nearby Galaxies \citep[COMING;][]{Sorai2019} that simultaneously presents {\COtw}, \COth\ and \CeiO\ J=1-0 maps observed with 45 m telescope of the Nobeyama Radio Observatory (NRO45M), and the EMPIRE survey and follow-up programs \citep{Cormier2018} that provide $^{12}\rm CO(1-0)$ and $^{13}\rm CO(1-0)$ observations with IRAM 30-m telescope. Then we combine the two survey with the data of these three CO isotopologues in \citet{Tan2011,Nakajima2018}, and the our observations (introduced in Appendix~\ref{sec:data_pmo}) using the Purple Mountain Observatory (PMO) 13.7-m millimeter telescope located in Delingha, China. For the galaxies without  $^{13}\rm CO(1-0)$ observations, we also use $^{12}\rm CO(1-0)$ data from the Nobeyama CO mapping survey \citep{Kuno2007} to complement. In total, we summarize 22 galaxies in $^{13}\rm CO(1-0)$ analysis, and 24 galaxies in $^{12}\rm CO(1-0)$ sample as listed in Table~\ref{tab:source}.

\begin{deluxetable}{lllcc}
\centering
  \tablecaption{The Basic Properties of Galaxies in the MALATANG Sample.}
  \label{tab:source}
  \tablewidth{0pt}
	\tablehead{   \colhead{Source}  &   \colhead{R.A.} & \colhead{Decl.} & \colhead{D} & \colhead{References} \\
     \colhead{} & \colhead{(J2000)} & \colhead{(J2000)} & \colhead{(Mpc)} & \colhead{} 
     }
     \decimalcolnumbers
  \startdata
*NGC660  &       $01^{\rm h}43^{\rm m}02.\!^{\rm s}4$ &       $ 13^{\circ}38'42.\!{''}0$      &   14.1         &     1       \\     
NGC891 &       $02^{\rm h}22^{\rm m}33.\!^{\rm s}4$ &       $ 42^{\circ}20'57.\!{''}0$      &     9.1        &     1       \\     
NGC2146 &       $06^{\rm h}18^{\rm m}37.\!^{\rm s}7$ &       $ 78^{\circ}21'25.\!{''}0$      &    18.0      &     1       \\     
NGC2903 &       $09^{\rm h}32^{\rm m}10.\!^{\rm s}1$ &       $ 21^{\circ}30'03.\!{''}0$      &    9.2       &     1,2       \\     
*NGC3079 &       $10^{\rm h}01^{\rm m}57.\!^{\rm s}8$ &       $ 55^{\circ}40'47.\!{''}0$      &   20.6       &     1       \\     
*NGC3627 &       $11^{\rm h}20^{\rm m}14.\!^{\rm s}9$ &       $ 12^{\circ}59'30.\!{''}0$      &   10.7       &     1,2       \\     
NGC3628 &       $11^{\rm h}20^{\rm m}17.\!^{\rm s}0$ &       $ 13^{\circ}35'23.\!{''}0$      &    11.3      &     1       \\     
NGC628 &       $01^{\rm h}36^{\rm m}41.\!^{\rm s}8$ &       $ 15^{\circ}47'00.\!{''}0$      &     8.6        &     1,2       \\     
NGC253 &       $00^{\rm h}47^{\rm m}33.\!^{\rm s}1$ &       $-25^{\circ}17'18.\!{''}0$      &     3.5      &     5,6     \\     
*NGC1097 &       $02^{\rm h}46^{\rm m}19.\!^{\rm s}0$ &       $-30^{\circ}16'30.\!{''}0$      &   15.4       &      $\dots$    \\
*NGC1365 &       $03^{\rm h}33^{\rm m}36.\!^{\rm s}4$ &       $-36^{\circ}08'25.\!{''}0$      &   18.1       &     $\dots$    \\
*NGC1808 &       $05^{\rm h}07^{\rm m}42.\!^{\rm s}3$ &       $-37^{\circ}30'47.\!{''}0$      &   9.5        &    $\dots$    \\
NGC4631 &       $12^{\rm h}42^{\rm m}08.\!^{\rm s}0$ &       $ 32^{\circ}32'29.\!{''}0$      &    7.5       &      3        \\     
NGC3184 &       $10^{\rm h}18^{\rm m}17.\!^{\rm s}0$ &       $ 41^{\circ}25'28.\!{''}0$      &    9.7         &      2        \\     
NGC5055 &       $13^{\rm h}15^{\rm m}49.\!^{\rm s}3$ &       $ 42^{\circ}01'45.\!{''}0$      &    8.8       &      2        \\     
NGC3521 &       $11^{\rm h}05^{\rm m}48.\!^{\rm s}6$ &       $ 00^{\circ}02'09.\!{''}0$      &    12.1      &      1        \\     
Maffei2 &       $02^{\rm h}41^{\rm m}55.\!^{\rm s}0$ &       $ 59^{\circ}36'15.\!{''}0$      &    3.5       &      6        \\     
*NGC1068 &       $02^{\rm h}42^{\rm m}40.\!^{\rm s}7$ &       $-00^{\circ}00'48.\!{''}0$      &   10.1       &      5,6      \\     
IC342   &       $03^{\rm h}46^{\rm m}48.\!^{\rm s}5$ &       $ 68^{\circ}05'47.\!{''}0$      &    3.4       &      5,6      \\     
M82     &       $09^{\rm h}55^{\rm m}52.\!^{\rm s}7$ &       $ 69^{\circ}40'46.\!{''}0$      &    3.5       &      1,3      \\     
Arp299 &       $11^{\rm h}28^{\rm m}30.\!^{\rm s}4$ &       $ 58^{\circ}34'10.\!{''}0$      &     54.1     &      $\dots$    \\
*NGC4736 &       $12^{\rm h}50^{\rm m}53.\!^{\rm s}0$ &       $ 41^{\circ}07'14.\!{''}0$      &   4.6        &      4,6      \\     
M51     &       $13^{\rm h}29^{\rm m}52.\!^{\rm s}7$ &       $ 47^{\circ}11'43.\!{''}0$      &    8.6       &      3,4,6    \\
M83     &       $13^{\rm h}37^{\rm m}00.\!^{\rm s}9$ &       $-29^{\circ}51'56.\!{''}0$      &    4.7       &       6   \\ 
NGC5457 &       $14^{\rm h}03^{\rm m}12.\!^{\rm s}5$ &       $ 54^{\circ}20'56.\!{''}0$      &    6.6       &       4,6  \\
NGC6946 &       $20^{\rm h}34^{\rm m}52.\!^{\rm s}3$ &       $ 60^{\circ}09'14.\!{''}0$      &    4.5       &       2   \\ 
NGC4254 &       $12^{\rm h}18^{\rm m}50.\!^{\rm s}0$ &       $ 14^{\circ}24'59.\!{''}0$      &    13.9      &       2   \\ 
NGC4321 &       $12^{\rm h}22^{\rm m}55.\!^{\rm s}0$ &       $ 15^{\circ}49'19.\!{''}0$      &    13.9      &       2   \\ 
\enddata
\tablecomments{ \\
(1): Galaxy name. Those marked with an asterisk (*) are AGN hosts. \\
(2) and (3): Galaxy center coordinates. \\
(4): Luminosity distance are the latest measurements that can be found in the NED database \citep{https://doi.org/10.26132/ned1} by 14nd, Spe, 2022. References: \citet{Nasonova2011,Radburn-Smith2011,Adamo2012,Tully2016,Tully2013,Sorce2014,Sabbi2018,Jang2018,Monachesi2016,Pejcha2015,Vacca2015,McQuinn2017,Hoeflich2017}.
(5): CO data sources, 1. the COMING survey \citealp{Sorai2019}; 2. EMPIRE survey, \citealp{Cormier2018}; 3. \citealp{Tan2011}; 4. this work; 5. \citealp{Nakajima2018}; and 6. \citealp{Kuno2007} (only \COtw). Note that \citet{Sliwa2012} provide high-resolution maps of $^{12}\mathrm{CO} (3-2)$, $^{12}\mathrm{CO} (2-1)$ and $^{13}\mathrm{CO} (2-1)$ observed  using the Submillimeter Array for ARP299, but are not included in this work to avoid uncertainty in the line ratio.
}
\end{deluxetable}

For the CO datacubes, we reproject them with $1/\sqrt{2} \times$ beam size to avoid over-sampling, and then extract the spectra for each new position. 
Based on these spectra that are converted in main-beam temperature ($T_{\rm mb}$) scales, after determining the velocity range of the line emission (from \COtw\ spectrum), we measure the new velocity-integrated CO line intensity in the velocity range, $I_{\rm CO} \equiv \int {T_{\rm mb}}\;dv$, and the new rms noise over the full spectrum except the emission line and edge channels. Then we estimate the uncertainty of the integrated intensity using the formula in \citet{Gao1996}: 
\begin{equation}
\label{rms_co}
  \Delta I_{\rm CO} \equiv \frac{T_{\rm rms}\Delta v_{\rm FWZI}}{\sqrt{f\left(1- \frac{\Delta v_{\rm FWZI}}{W}\right)}}, 
\end{equation} 
where $f \equiv \Delta v_{\rm FWZI} /\delta v $, $\Delta v_{\rm FWZI}$ is the full width at zero intensity (FWZI) of the emission feature, $\delta v $ is the velocity channel width,  and $W$ is the entire
velocity coverage of the spectrum. 
We treat the positions with a signal to noise ratio $S/N \geq$ 3 as detection, and give an (valuable) upper limit of 3 times the uncertainty for the other ones (tentative detection $S/N \geq$ 1.5).
Finally, we derived the luminosities of \COtw, \COth\ and \CeiO, and the molecular gas column density from the velocity-integrated CO line intensity $I_{\rm CO}$ using an empirical $X_{\rm CO} = 2\times 10^{20}$ cm$^{-2}$ (K km s$^{-1})^{-1}$ \citep{Nishiyama2001} corresponding to the galactic conversion factor $\alpha_{\rm CO} = 3.2$ \msolar (K km s$^{-1} {\rm pc^{2}})^{-1}$.

Assuming that the molecular cloud is under local thermal equilibrium (LTE) conditions and \COth\ has the same excitation temperature as \COtw, for the pixels with both \COtw\ and \COth\ detections, we can calculate the \COth\ optical depth averaged in the beam from

\begin{equation}
  \tau (I_{\rm^{13} CO}) \equiv {\rm ln}\left[1-{\frac{\int T_{\rm mb}\left({\rm^{13} CO}\right) \;dv}{\int T_{\rm mb}\left({\rm^{12} CO}\right) \;dv}}\right]^{-1},
\end{equation}
where $T_{\rm mb}$ is equivalent to the ${T^{*}_{\rm R}}$ (radiation temperature) in the notation of \citet{Kutner1981}.
Then we use the definition of $N{\rm (H_{2})(^{13}CO)}$ in \citet{Wilson2009}, which assume  the \COth\ abundance ${\rm [^{13}CO]/[H_{2}]}$ is $8\times\ 10^{-5}/60$ \citep{Frerking1982}, and equate the column densities derived from both \COtw\ and \COth, we can estimate the kinetic temperature of the gas, $T_{\rm K}$.

\subsection{\textit{WISE} 12 \micron\ Data}
We downloaded the 2 deg $\times$ 2 deg assembled \textit{WISE} Atlas tiles (image and uncertainty files) for each source from the NASA/IPAC Infrared Science Archive \citep{https://doi.org/10.26131/irsa153}. Based on them, we produced the background map using the SExtractor package \citep{Bertin1996}. After masking some saturated pixels (mainly for M82 and NGC1068) and subtracting the background, we smoothed the flux images using the convolution kernels provided by \citet{Aniano2011} to mach the beam size of CO or other gas traces. Combining the smoothed flux images and uncertainty images, we can get the instrumental uncertainty, and add it with the zero-point uncertainty in quadrature to get the total uncertainty of 12 \micron\ flux in corresponding pixel. The recovered core of saturation were not used due to the large uncertainty (20\% to 30\%) as measured by \citet{Jarrett2019}, which lead to, in this work, missing W3 data corresponding to the CO data of NGC1068 provided by \citet{Nakajima2018}. 
We can convert the maps in original unit (digital numbers, DN) to the luminosity maps in units of $L_\odot$, following the formula:
\begin{equation}
\label{lum_w3}
  \Big(\frac{L_{12 \micron}}{L_\odot}\Big) \equiv 7.042 \times\ \Big(\frac{F_{12 \micron}}{\rm DN}\Big) \Big(\frac{D_L}{\rm Mpc}\Big)^2   ,
\end{equation}
which is computed using the W3 zero-point magnitude MAGZP = 18.0 mag, the isophotal flux density $S_0$ = 31.674 Jy, and the bandwidth $\Delta \nu = 1.1327 \times 10^{13}$ Hz \citep{Jarrett2011}. 
Then scale the value by a factor of 1.133 $\times$ (beam size/pixel size)$^2$ to get the 12 \micron\ luminosity \Lwc\ correspond to gas traces L$_{\rm gas}$. Some equations and parameters in calculating the monochromatic luminosity and uncertainty at \textit{WISE} 12 \micron\ band are described in detail in \citet{Chown2021}.

\subsection{JCMT HCN(4-3) and HCO$^{+}$(4-3) Data}

The dense gas data are from the MALATANG survey, which aims to construct a largest and deepest HCN(4-3) and HCO$^{+}$(4-3) maps with FWHM angular resolution about $14''$ in a flux-limited sample of nearby IR-bright galaxies. \citet{Tan2018} present the first results of six galaxies, NGC253, NGC1068, IC342, M82, M83, and NGC6946, that were mapped in their central $2' \times 2'$ regions using a 3 $\times$ 3 jiggle mode (i.e. rapidly step the secondary mirror to observe a grid-pattern of 9 points) with grid spacing of 10 arcsec. Our analysis about dense gas tracers is limited to the maps of these six galaxies.

After producing the pipeline-processed spectra of HCN(4-3) and HCO$^{+}$(4-3) using STARLINK software package \citep{Currie2014} and the ORAC-DR pipeline \citep{Jenness2015}, we flag some sub-scans with unstable baselines or abnormal noise levels as bad quality using the {\tt CLASS} package \citep{Pety2005}, and then we can get average spectrum for each position. We align the ancillary CO data \citep{Kuno2007,Wilson2012} with the datacube to determine the velocity range over which CO emission line is significant, and then fit and subtract a first-order baseline using the rest channels. 
Based on the final spectra, we also use CO-emitting velocity ranges to measure the integrated-intensity of the dense gas tracers. Then we calculate the luminosities of dense gas tracers for all detected positions following the formula in \citet{Solomon1997}: 
\begin{equation} 
\begin{split}
\label{dens_lum}
L'_{\mathrm{dense}} = 3.25 \times 10^{7}\Big(\frac{S\Delta v}{\rm 1 ~Jy~km~s^{-1}} \Big) \Big(\frac{\upsilon_{\rm obs}}{\rm 1~GHz} \Big)^{-2}\\ \times~\Big(\frac{D_{\rm L}}{\rm 1~Mpc} \Big)^{2} (1 + z)^{-3}~{\rm K~km~s^{-1}~pc^{2}}.
\end{split}
\end{equation}
One notable point is that this work include all pixels with $S/N \geq$ 3 even in the outer disks (i.e. $\geq 1'$), which (13 pixels) are not shown in the table provided by \citet{Tan2018}.

\section{Correlations between molecular gas tracers and \textit{WISE} 12 \micron\ luminosities}
\label{sec:gas_w3}

In this section, we first revisit the correlation between 12 \micron\ and $^{12}\rm CO (1-0)$ luminosity on kpc or sub-kpc scales for as much pixels as possible in the MALATANG sample. We then extend the relation to other emission lines including the optically thin isotopologues $^{13}\rm CO$ and dense molecular tracers. By comparing these relations over the same set of pixels, we explore the tightest correlation and the physical reason behind the correlations between 12 \micron\ emission and molecular gas.

\subsection{The $L_{\rm CO(1-0)}-L_{\mathrm{12\mu m}}$ Relation}

\begin{figure*}[!t]
\centering \includegraphics[width=0.31\textwidth]{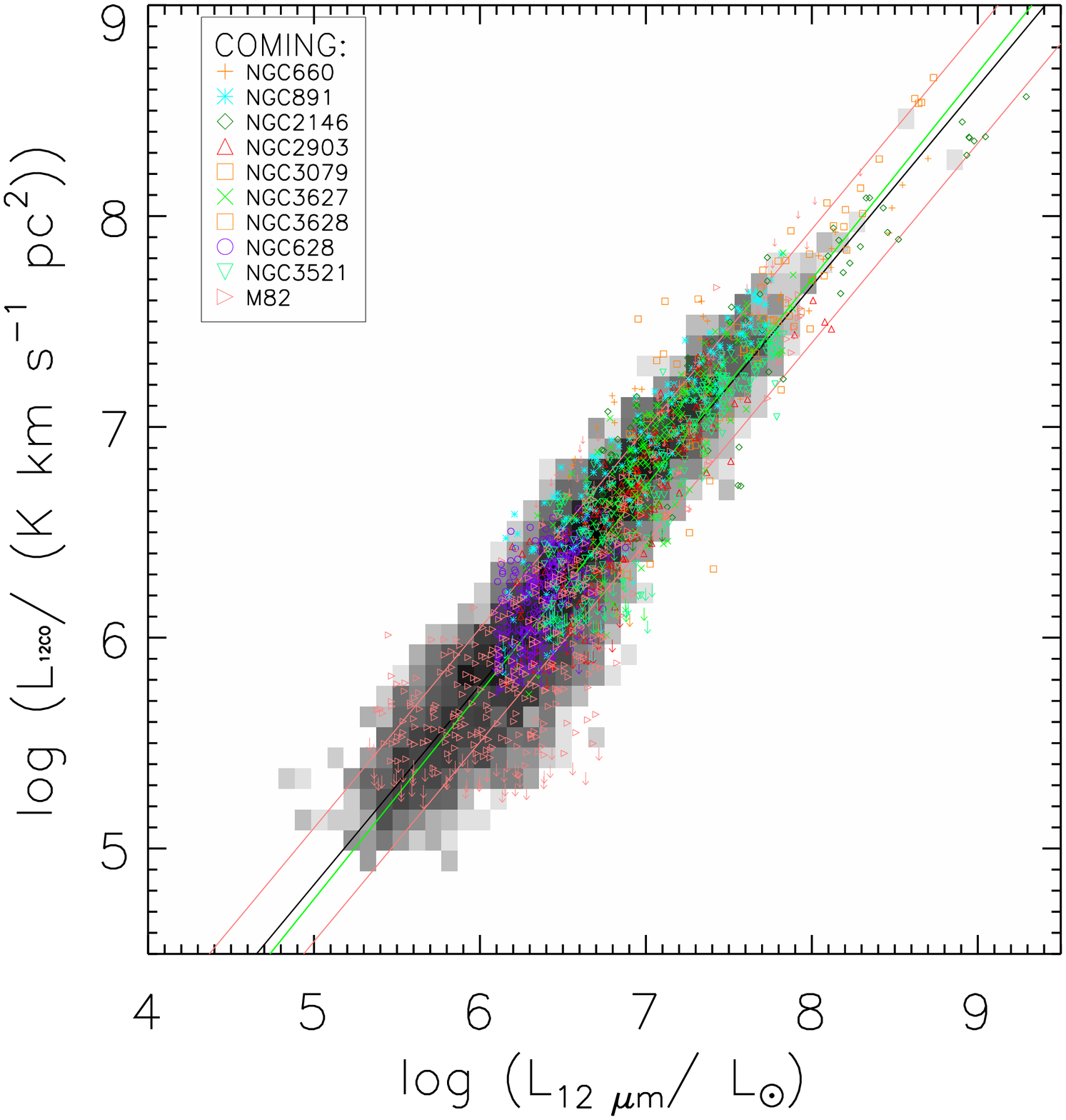}
\centering \includegraphics[width=0.31\textwidth]{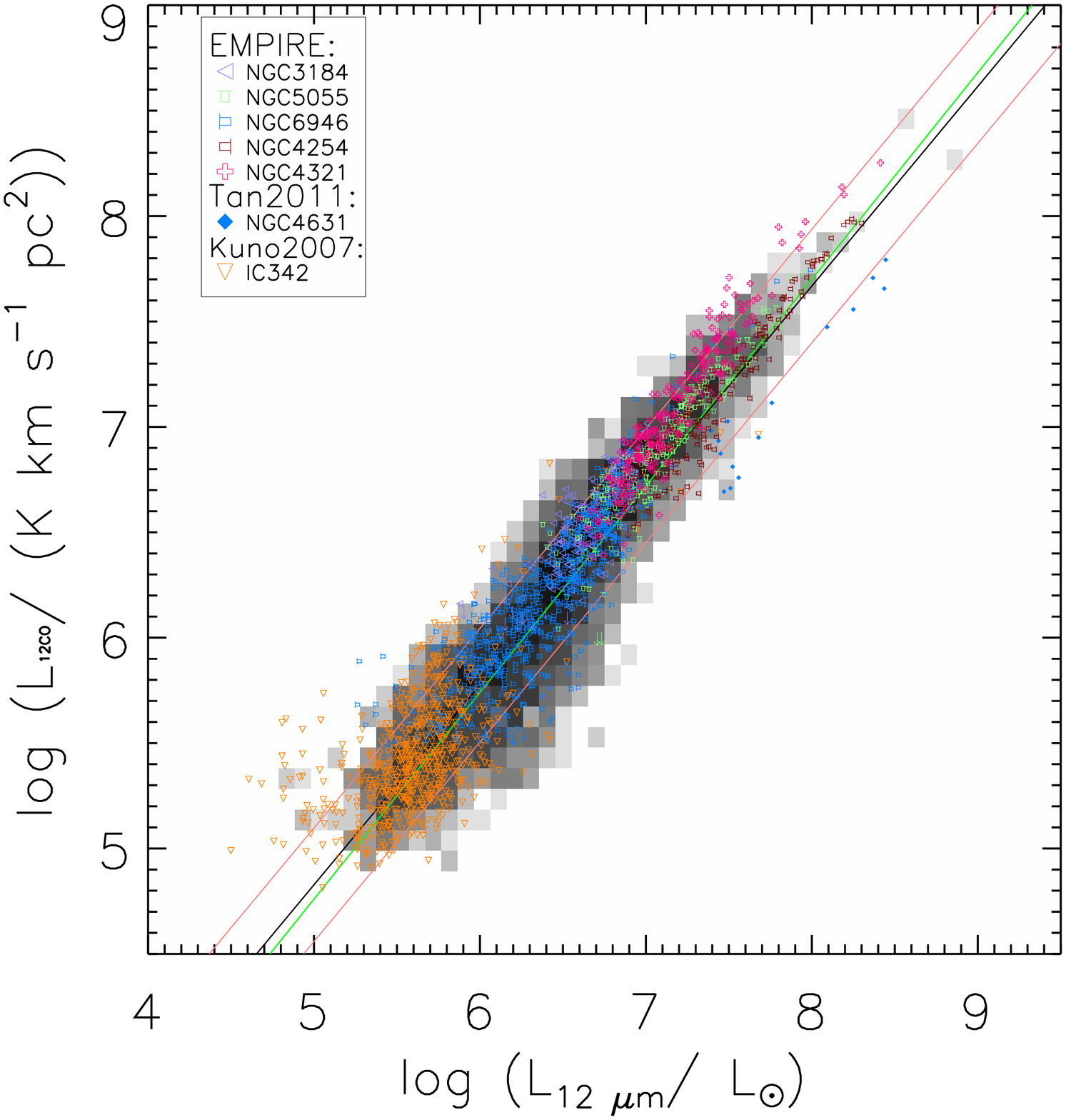}
\centering \includegraphics[width=0.372\textwidth]{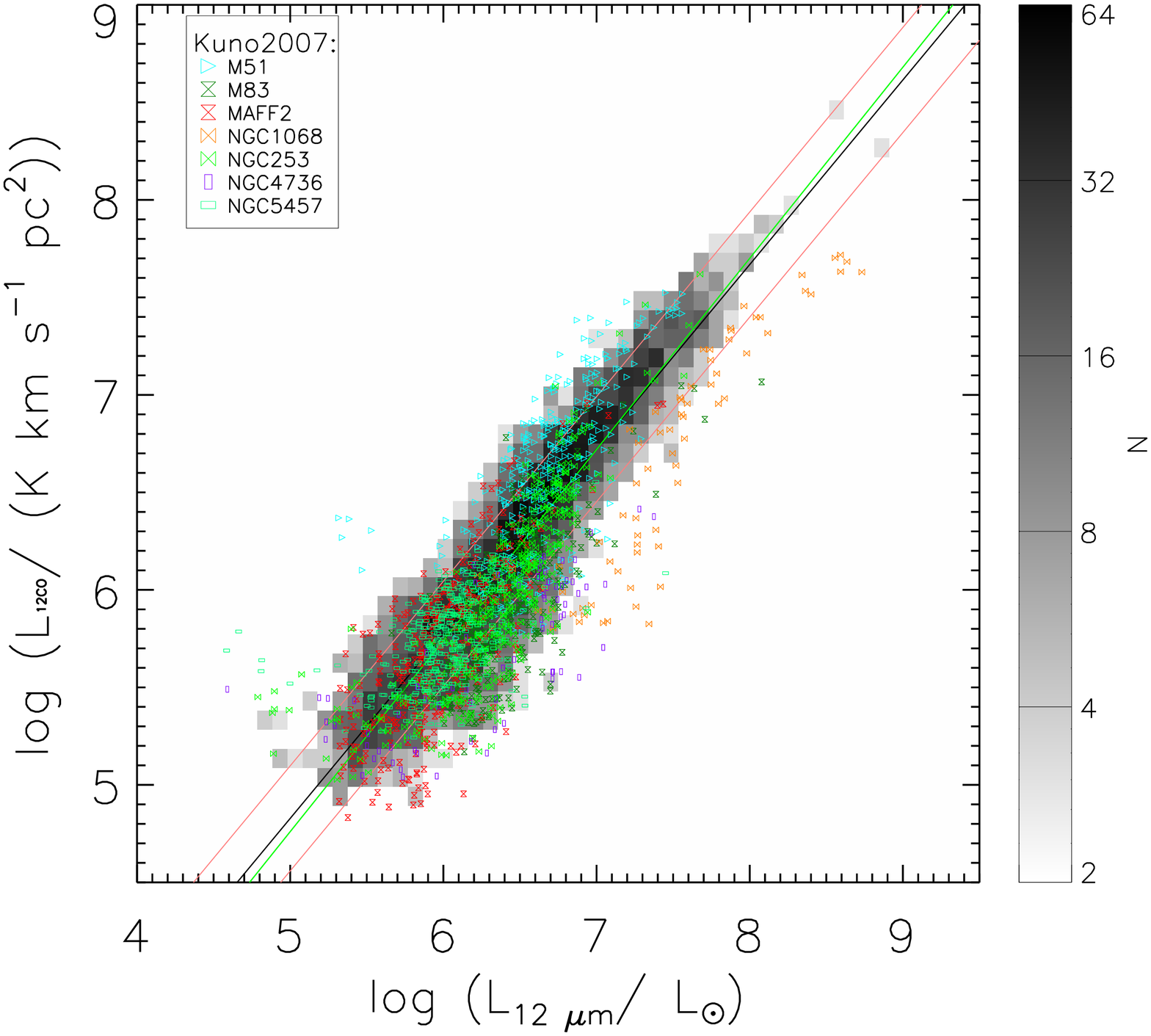}
\caption{
Correlation of the {$^{12}\rm CO$} luminosities with the 
  mid-infrared monochromatic luminosities ($L_{12 \micron}$) from \textit{WISE} measured in the 12 band for galaxies spatially resolved on kpc or sub-kpc scales. CO detections from different galaxies galaxy are indicated with different symbols and colours as indicated in the upper left corner, while a few valuable upper limits are also plotted as downward arrows. The galaxies are divided into three sub-samples in these three panels, and all detected pixels in the entire sample are plotted as a gray-scale background for comparison. 
  The black line and two dashed red lines depict respectively the best-fitting relation (with parameters listed in
  Table~\ref{tab:tbl1}) and the 1 $\sigma$ total/observed scatter, while the green line show the relation of global galaxies from \citet{Gao2019}. The median measurement uncertainties are illustrated by the characteristic error bars shown in the lower right corner.} 
\label{fig:co_w}
\end{figure*}
Figure~\ref{fig:co_w} shows the correlation of the $^{12}$CO(1-0) luminosities, $L_{\rm ^{12}CO(1-0)}$ and mid-infrared luminosities measured from \textit{WISE} images in the 12 \micron\ band at small scales of $\sim$ 0.2-1.3 kpc for the 24 MALATANG galaxies (indicated with different symbols and colors). For some galaxies with available \COtw\ data from more than one reference (as shown in Table~\ref{tab:source}), we just use data from the one that can provide most pixel to compute the pixel measurements of log$L_{\rm ^{12}CO}$(y-axis). Distributions of all detections are shown as a gray-scale background in both panels. The average uncertainty of ${\rm ^{12}CO(1-0)}$ luminosities measured using Eq.~\ref{rms_co} in all pixels is about 0.045 dex, and the one of 12 \micron\ is 0.021 dex (which includes both the photometric uncertainty and the uncertainty of the magnitude zero-point measured by \citealp{Jarrett2011}), displayed as the error bar in the lower right corner. 

Then we use the IDL script \textit{LinMix} \footnote{Available from the NASA IDL Astronomy User's Library
  \url{https://idlastro.gsfc.nasa.gov/ftp/pro/math/linmix_err.pro}} \citep{Kelly2007}  which is a Bayesian linear regression taking into account upper limits and uncertainties in both the x- and y-axes. Fitting $L_{\rm ^{12}CO}$ in ${\rm K\;km\;s}^{-1}\;{\rm pc^{2}}$ versus $L_{12 \micron}$ in $L_\odot$
  yields the best-fit result
\begin{equation} 
\label{co_12}
\log L_{\rm ^{12}CO} = (0.95 \pm 0.01) \log L_{12 \micron}+(0.09 \pm 0.04).
\end{equation}

 We can see the best fitting of the entire MALATANG sample is almost the same as the result (the green line) of global galaxies performed by \citet{Gao2019}, and show slightly larger total/observed scatter ($\sigma_{\rm tot} = 0.27$) but smaller intrinsic scatter ($\sigma_{\rm int} = 0.07$). So based on the MALATANG sample, we can extend this relation to low-luminosity end (down to $10^5\;[{\rm K\;km\;s}^{-1}\;{\rm pc^{2}}]$ and $10^5$ \Lsolar\ in both $L_{\rm ^{12}CO(1-0)}$ and $L_{12 \micron}$), though the relation may be are various from galaxy to galaxy as indicated by \citet{Chown2021}.
 
\begin{figure*}[!t]
\centering \includegraphics[width=0.47\textwidth]{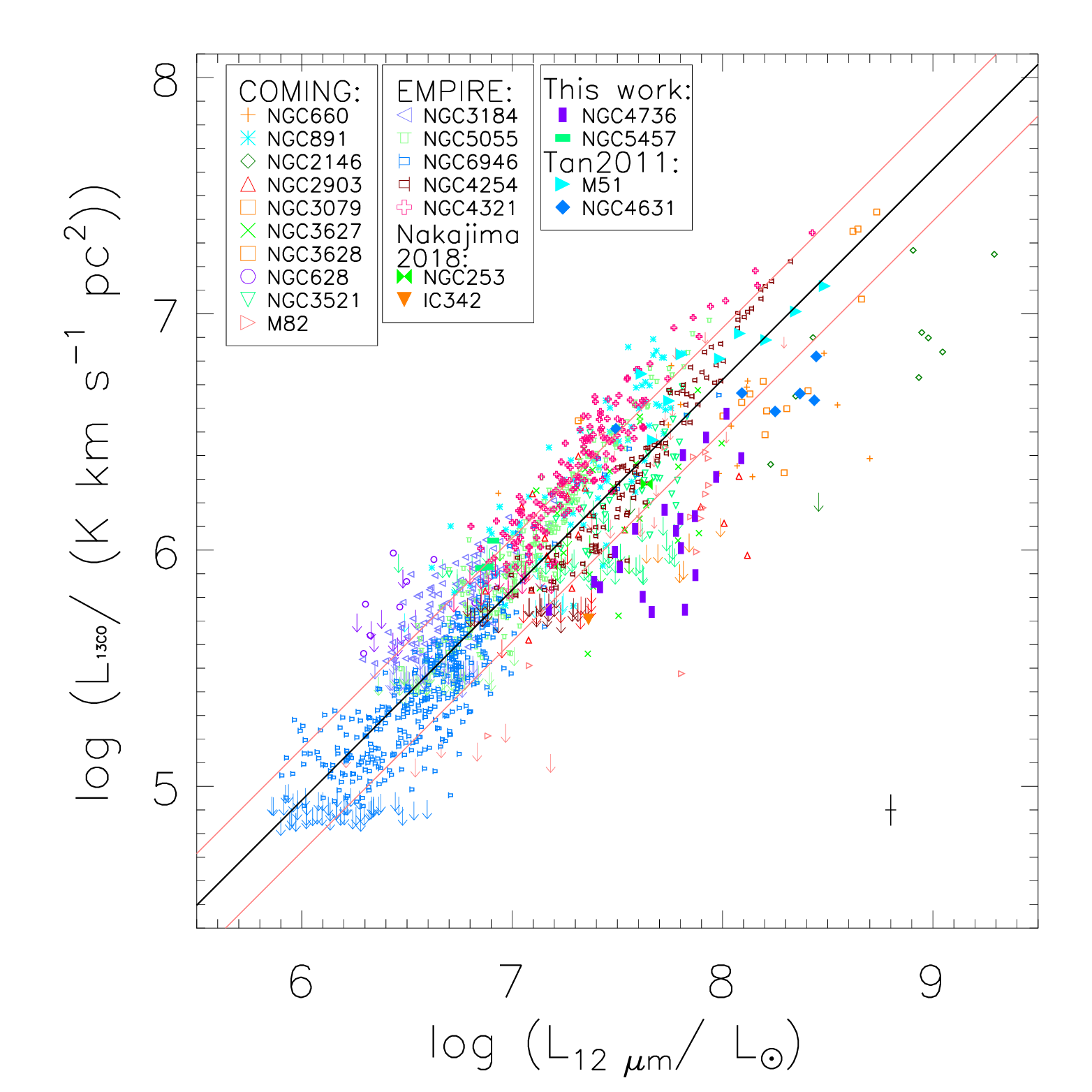}
\centering \includegraphics[width=0.47\textwidth]{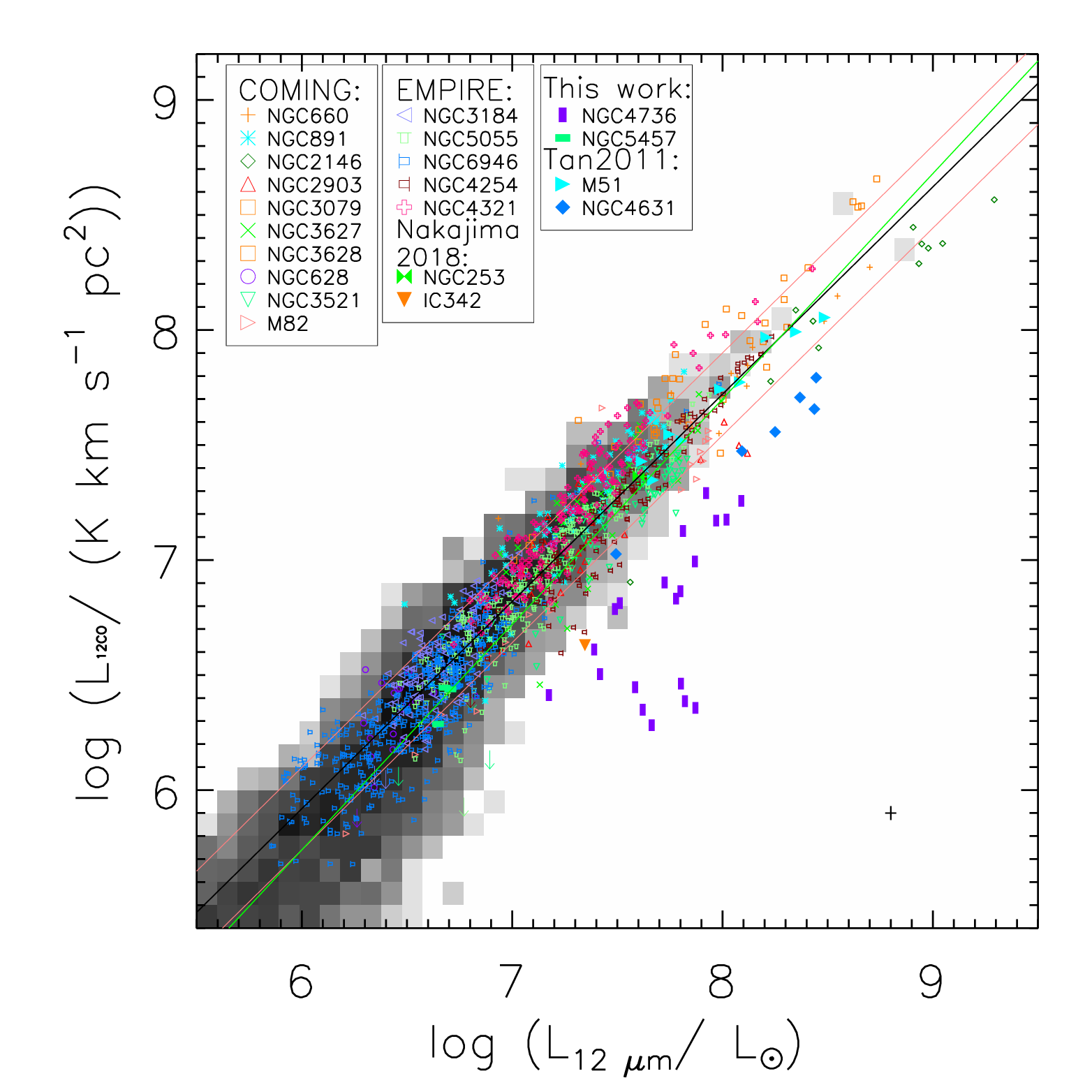}
\caption{Same as Figure~\ref{fig:co_w}, the relation between \COth\ (left panel) and \COtw\ (right panel) luminosities and ${12 \micron}$ luminosities, but only the pixels with significant {$^{13}\rm CO$} emission (detected or with valuable upper limits) are shown. Large filled symbols are used to indicate the galaxies observed by us with PMO and the center of galaxies observed by \citet{Nakajima2018}. The parameters of best-fitting linear relations are also listed in Table~\ref{tab:tbl1}. And in the right panel, the gray-scale background in Figure~\ref{fig:co_w} is also plotted for comparison. } 
\label{fig:co2_w}
\end{figure*}

 Analogously, we plot the similar relation between \COth\ and 12 \micron\ luminosities based on $\sim$ 1000 pixels (with \COth\ detections or valuable upper limits) in the left panel of Figure~\ref{fig:co2_w}, and get the best-fitting linear relation
\begin{equation} 
\label{co_13}
\log L_{\rm ^{13}CO} = (0.89 \pm 0.01) \log L_{12 \micron}-(0.39 \pm 0.08).
\end{equation}

For comparison, we also show the relation of \COtw\ based on the same pixels in the right panel. The points with same colour and symbol in these two panel belong same galaxy as in Figure~\ref{fig:co_w}, and the large filled symbols in Figure~\ref{fig:co2_w} are 6 galaxies observed in \COtw\ and \COth\ simultaneously by us with PMO and provided by \citet{Nakajima2018}. We can see there are strong positive correlations in both panels, though a certain fraction of \COth\ data are upper limits. But the total and intrinsic scatter ($\sigma_{\rm tot} = 0.21$ and $\sigma_{\rm int} = 0.05$) of \COth\ vs. 12 \micron\ relation is larger than the one of \COtw\ ($\sigma_{\rm tot} = 0.18$ and $\sigma_{\rm int} = 0.02$), meanwhile the Spearman’s correlation coefficient ($r$) also is a little smaller (0.90 compared to 0.94). The relations between \COth\ and 12 \micron\ are more remarkably varied in different galaxies (discussed in Sec~\ref{sec:diff_sample}). These results suggest the 12 \micron\ luminosities are more tightly correlated with the \COtw\ luminosities than with \COth\ luminosities over the same set of pixels. 
In Sec~\ref{sec:offset_gas}, we will try to figure out the parameter that could contribute the scatter of the relation to confirm the application of these estimator.

 Just using \COth\ pixel sample to perform fitting in the right panel of Figure~\ref{fig:co2_w}, besides with decreased scatter and increased correlation coefficient, we can find the new best-fitted \COtw\ vs. 12 \micron\ relation yields a little higher \COtw\ luminosity and a shallower slope than the global one (the green line). This change can be simply and naturally caused by selection effect. There are two kinds of effect: the missing position with undetected/faint \COth\ emission are more likely to have lower \COtw\ luminosity, which lead to the relation be higher, and the missing fraction could increase with reduced 12 \micron\ luminosity (correlated with CO luminosity), which make the slope lower and difference larger and more significant towards the low end.
  In the discussion that follows (Sec~\ref{sec:diff_sample}), we will further analyse the effect of sample selection.

 \begin{deluxetable*}{ccccccccc}
        \centering
	\tablecaption{Best-fit relations between (dense) gas tracers and  W3 (12 \micron) luminosities.}
	\label{tab:tbl1}
	\tablewidth{0pt}
     \tablehead{
 \colhead{Gas tracer} &	\multicolumn{2}{c}{Number of pixels}	&	\colhead{$k$}			&	\colhead{$b$}	&	\multicolumn{2}{c}{Scatter}	&	\colhead{$r$}		&	 \colhead{Figure}	\\
   \colhead{} & \colhead{Detections} & \colhead{Upper limits}	& \colhead{} 	& \colhead{}	& \colhead{$\sigma_{\rm tot}$}	 & \colhead{$\sigma_{\rm int}$} & \colhead{}	& \colhead{}	 	 
   }
  \decimalcolnumbers
  \startdata
\COtw &	5660 & 292  	&	0.95	$\pm$	0.01	&	$0.09	\pm	0.04$ &	0.27 & 0.07	&	 0.92	&  Figure~\ref{fig:co_w}	\\
\COth	&	 1179 & 277	&	0.89 	$\pm$	0.01	&	$-0.39	\pm	0.08$	&	0.21 & 0.05	&	 0.90 $\pm$	0.01	&  Left panel, Figure~\ref{fig:co2_w}	\\
\COtw\ (matched w/ \COth)	&	1448 & 8	&	0.90	$\pm$	0.01	&	0.51	$\pm$	0.06 &	0.18 & 0.02	&	 0.94	&  Right panel, Figure~\ref{fig:co2_w}	\\
HCN(4-3)	&	81 & 0	&	0.42	$\pm$	0.04	&	2.40	$\pm$	0.31 &	0.29 & 0.09	&	 0.77 $\pm$	0.05	&  Left panel, Figure~\ref{fig:dens_w}	\\
HCO$^{+}$(4-3)	&	76 & 0	&	0.47	$\pm$	0.06	&	2.06	$\pm$	0.43 &	0.33 & 0.11	&	 0.73 $\pm$	0.07	&  Middle panel, Figure~\ref{fig:dens_w}	\\
\COtw\ (matched w/ HCN(4-3))	&	81 & 0	&	0.70	$\pm$	0.04	&	1.82	$\pm$	0.27 &	0.26 & 0.06	&	 0.92 $\pm$	0.02	&  Right panel, Figure~\ref{fig:dens_w}\\
\CeiO\ &	37 & 0 	&	0.55	$\pm$	0.10	&	1.57	$\pm$	0.70 &	0.33 & 0.11	&	 0.71 $\pm$	0.10	&  Figure~\ref{fig:co_w} \\
  \enddata
\tablecomments{The relations are parameterized as $y = k x + b$, with all the quantities given in this table, where y is log $L_{\rm Gas~tracer}~[\mathrm{K~km~s^{-1}~pc^{-2}}]$ and x is $\log L_{\mbox{12\micron}}~[\mathrm{L_\odot}]$. The derived intrinsic scatter and Spearman's correlation coefficient of each relation is listed as $\sigma_{\rm int}$ and $r$, after taking into account upper limits and uncertainties. }
\end{deluxetable*}

\begin{figure*}[!t]
\centering \includegraphics[width=0.33\textwidth]{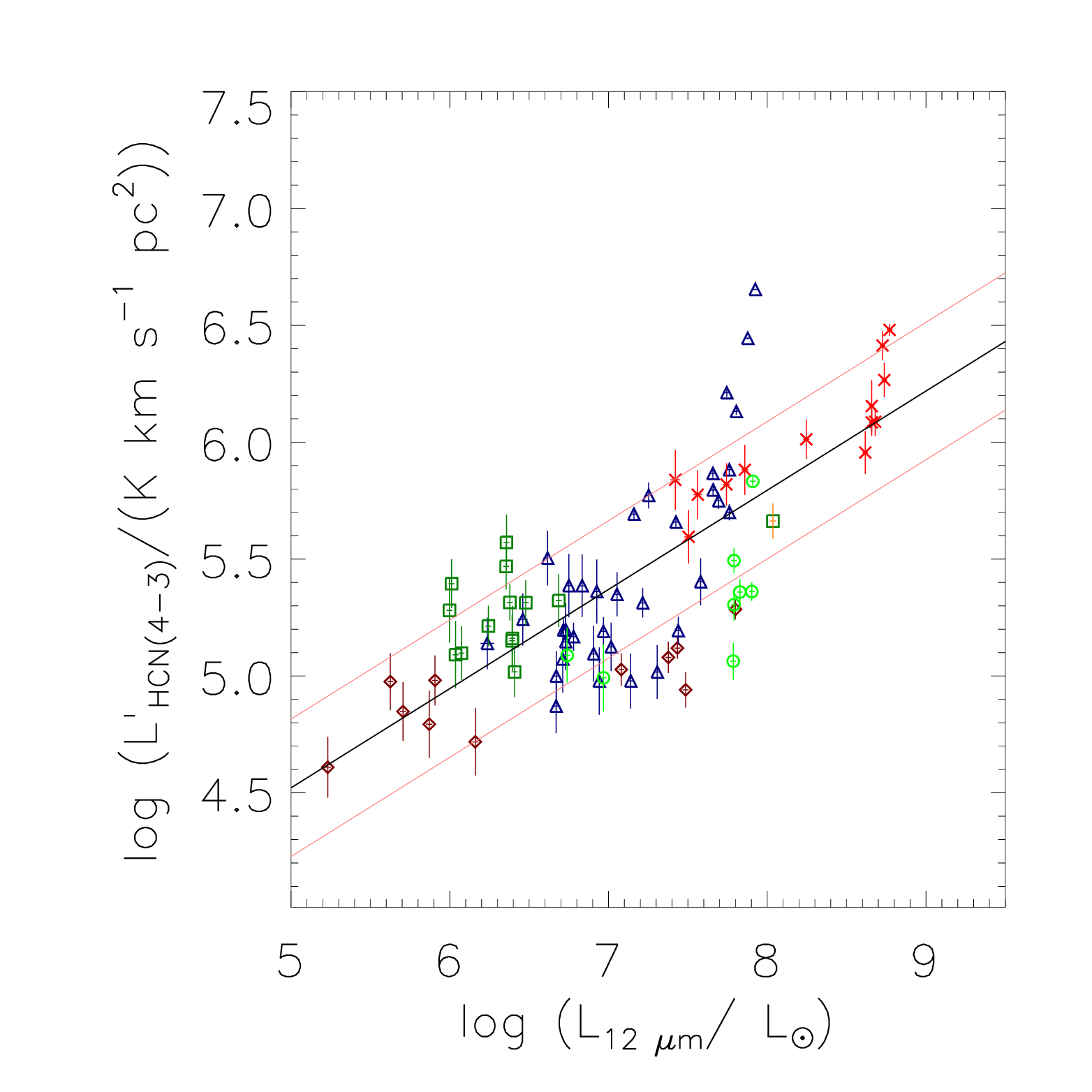}
\includegraphics[width=0.33\textwidth]{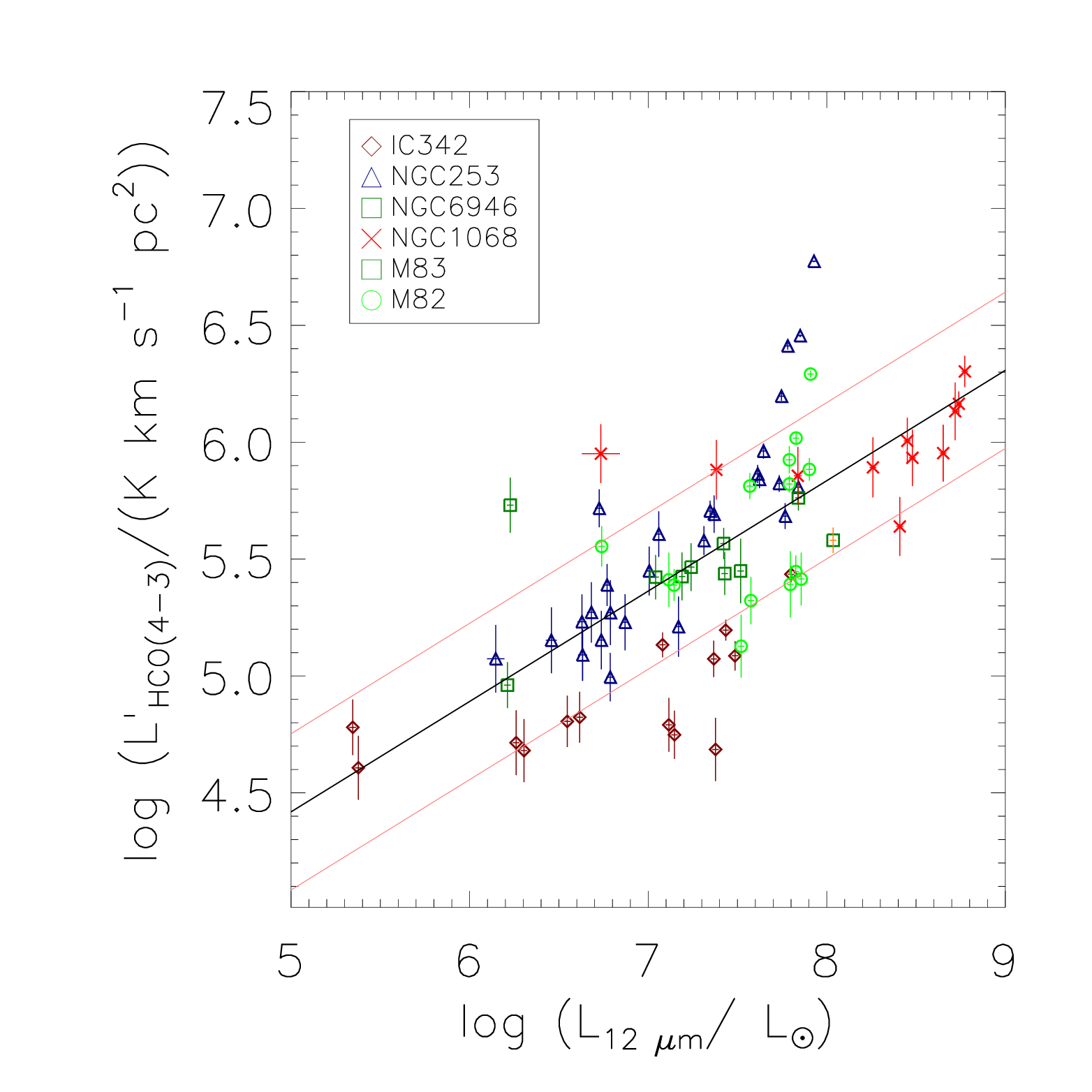}
\includegraphics[width=0.33\textwidth]{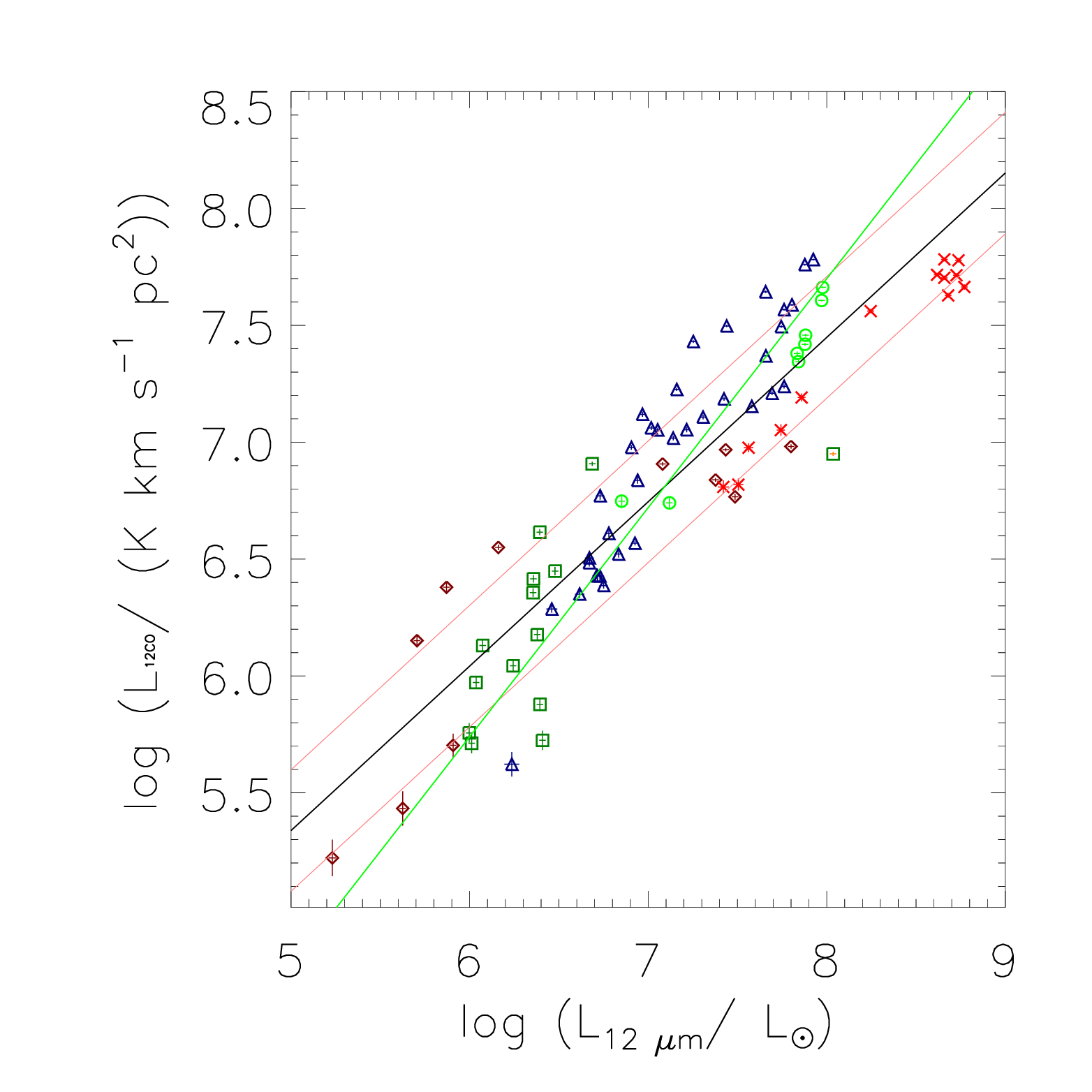}
          		
\caption{The relation between ${12 \micron}$ luminosities and Y. From left to right, Y is defined as the HCN(4-3) (detected) luminosities,  HCO$^{+}$(4-3) (detected) luminosities, and $^{12}\rm CO (1-0)$ luminosities for the same pixels of HCN detections. Different colours and symbols indicate detected pixel in different galaxies, and the error bars show their measurement uncertainties. In each panel, the black and two red lines show, respectively, the best-fitting linear relation (with parameters listed in Table~\ref{tab:tbl1}) and the 1$\sigma$ total/observed scatter, while the green line in the rightest panel still indicate the relation of global galaxies from \citet{Gao2019}.} 
\label{fig:dens_w}
\end{figure*}

\subsection{Correlations Between Dense Gas Tracers and 12\micron\ Luminosities}
\label{sec:dens_w3}
A main aim of this work is to figure out the origin of the correlation between 12 \micron\ and CO (molecular gas) emission. It is presumed that the correlation is like the KS relation, in other words, the 12 \micron\ emission (maybe dominated by 11.3 \micron\ PAH bands) is an accurate, quantitative measure of SFR \citep{Xie2019,Cluver2017}, so the correlation can be simply understood as the formation of stars from molecular gas. 
If this conjecture is right, the spatially resolved relation between the masses of dense molecular clumps (traced by HCN(4-3) and/or HCO$^{+}$(4-3) luminosity) and stars formed (traced by 12\micron\ luminosity) should be more linear and tight than the correlation with total molecular mass (traced by CO luminosity).

In Figure~\ref{fig:dens_w}, we show the ${L^{'}_{\rm dense}}$-${L_{\mbox{12\micron}}}$ relations, which are based on all HCN(4-3) or HCO$^{+}$(4-3) detected regions/pixels in MALATANG maps (different galaxy is indicated using different symbol and colour). The rightest panel shows the 12 \micron\ vs. $^{12}\rm CO (1-0)$ relation over the same set of HCN(4-3) detected pixels. In this panel, we derive the $^{12}\rm CO (1-0)$ luminosity of M82 from COMING survey \citep{Sorai2019}, and from the spatially resolved spectra provided by \citet{Kuno2007} for other galaxies, which are observed using NRO45M with similar angular resolution.

Then we can see the ${L^{'}_{\rm dense}}$-${L_{\mbox{12\micron}}}$ relations is nonlinear with larger scatter of 0.29 and 0.33 dex, compared with the $L_{\rm^{12} CO(1-0)}$-$L_{\mbox{12\micron}}$ relation. 
 We also compute the best linear least-squares fitting (logarithmic) and intrinsic scatter in the three scaling relations, with parameters shown in Table~\ref{tab:tbl1}.
The parameters also illustrate that $L_{\rm^{12} CO}$ is more tightly correlated with $L_{\mbox{12\micron}}$, with a very small intrinsic scatter of 0.06 dex, than dense gas (0.09 or 0.11 dex). Meanwhile, we also do the fits with ${{\rm log} L_{\mbox{12\micron}}}$ on the y-axis, the total (and intrinsic) scatters are 0.53, 0.50 and 0.34 (0.29, 0.25 and 0.12) dex for HCN(4-3), HCO$^{+}$(4-3) and \COtw, which are much larger than the ones of correlations shown in Figure~\ref{fig:dens_w}. Based on these results, we conclude that the relation between 12 \micron\ and $^{12}\rm CO (1-0)$  is unlikely to be due to star-formation law, though this sample is small.

We also notice there is one AGN (NGC1068) in these six nearby galaxies, whose data points are indicated as the red crosses in the panels.
If we remove this galaxy, the slope of $L_{\rm^{12} CO(1-0)}$-$L_{\mbox{12\micron}}$ will be much steeper (about 0.9) and in general consistent with the one in Figure~\ref{fig:co_w}, while the slope and correlation coefficients of ${L^{'}_{\rm HCN(4-3)}}$-${L_{\mbox{12\micron}}}$ and  ${L^{'}_{\rm HCO^{+} (4-3)}}$-${L_{\mbox{12\micron}}}$ relations become much lower and keep constant. 
So we speculate the relation of $L_{\rm^{12} CO(1-0)}$-$L_{\mbox{12\micron}}$ could be further complicated by the possible effects of AGN, which modify the expected bias in such small sample (detail discussion about AGN in Sec~\ref{sec:diff_agn}).

Based in these analysis, we conclude that the \textit{WISE} image can be used to derived molecular gas mass more directly and simply than using some kinds of star formation law, so the estimated molecular gas mass can be used widely, e.g. in studying the star formation process.

\subsection{Sources of Offsets from the Main Relation}
\label{sec:offset_gas}
About the correlations between $L_{\rm gas}$ and $L_{\mbox{12\micron}}$ shown above, the scatter and some systematic deviations are significant different.
We hereby examine the scatters in these relations as a function of the (spatially resolved) physical conditions of molecular gas.

\begin{figure*}[!t]
\centering \includegraphics[width=0.32\textwidth]{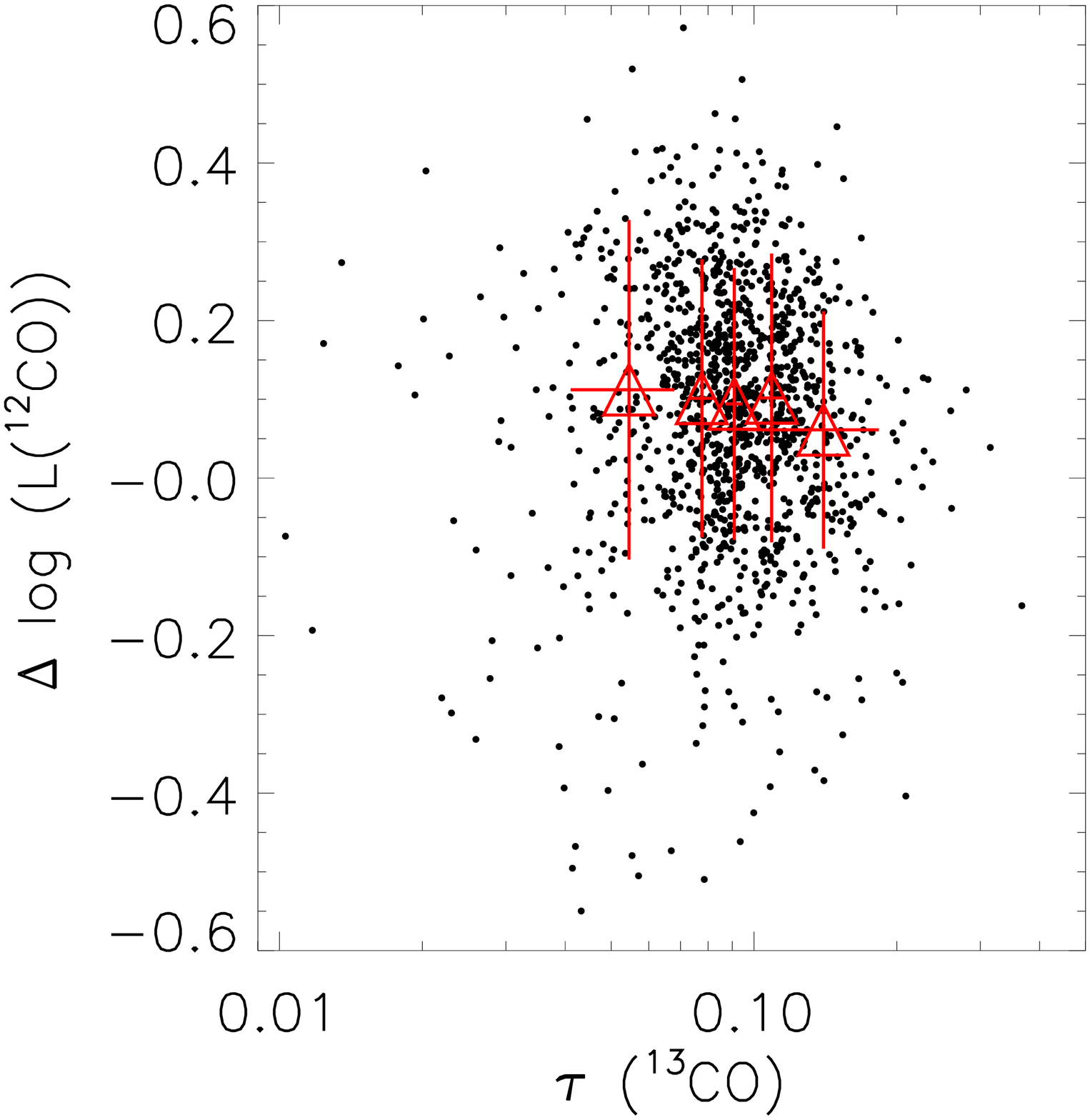}
\includegraphics[width=0.32\textwidth]{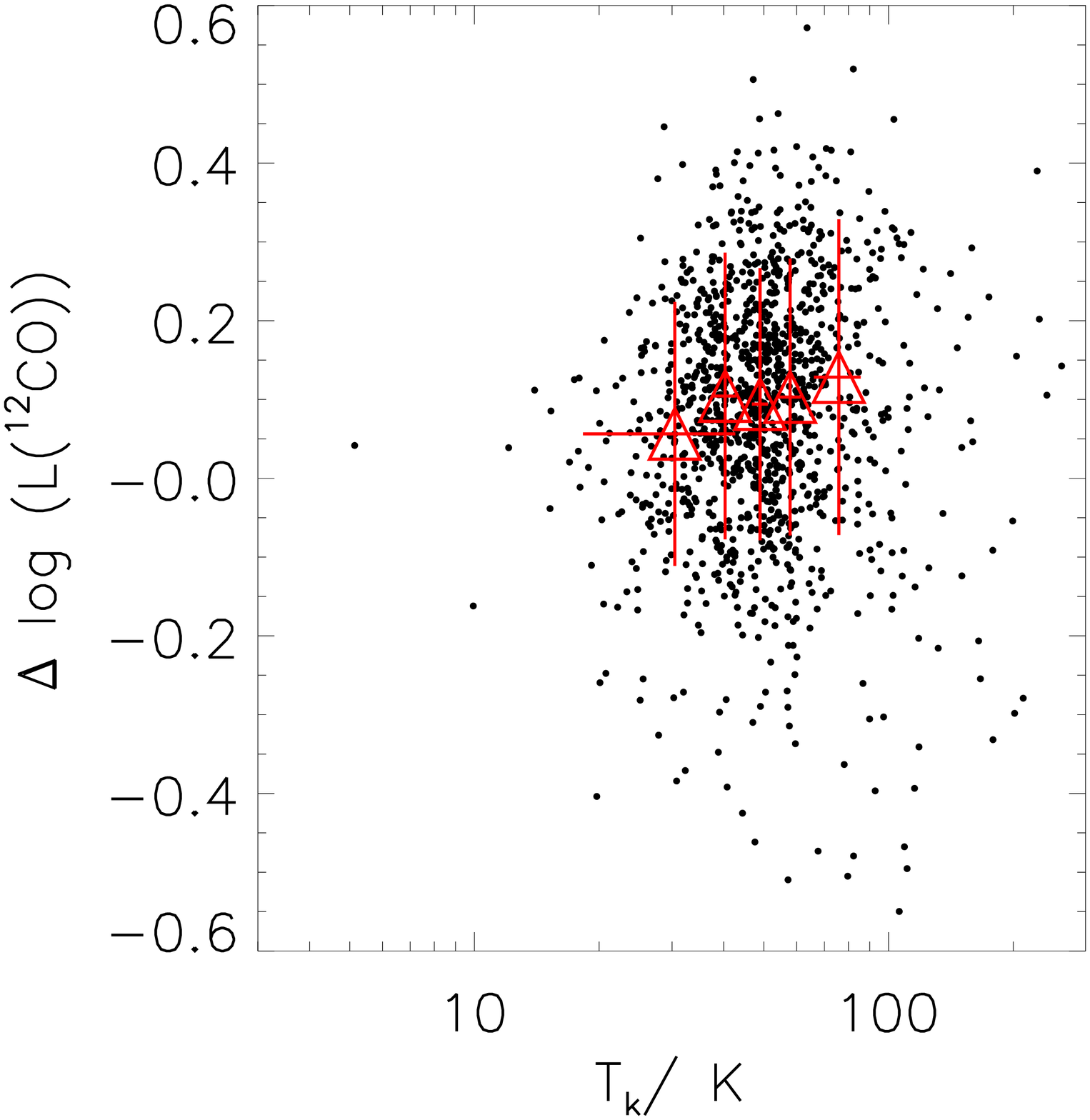}
\includegraphics[width=0.32\textwidth]{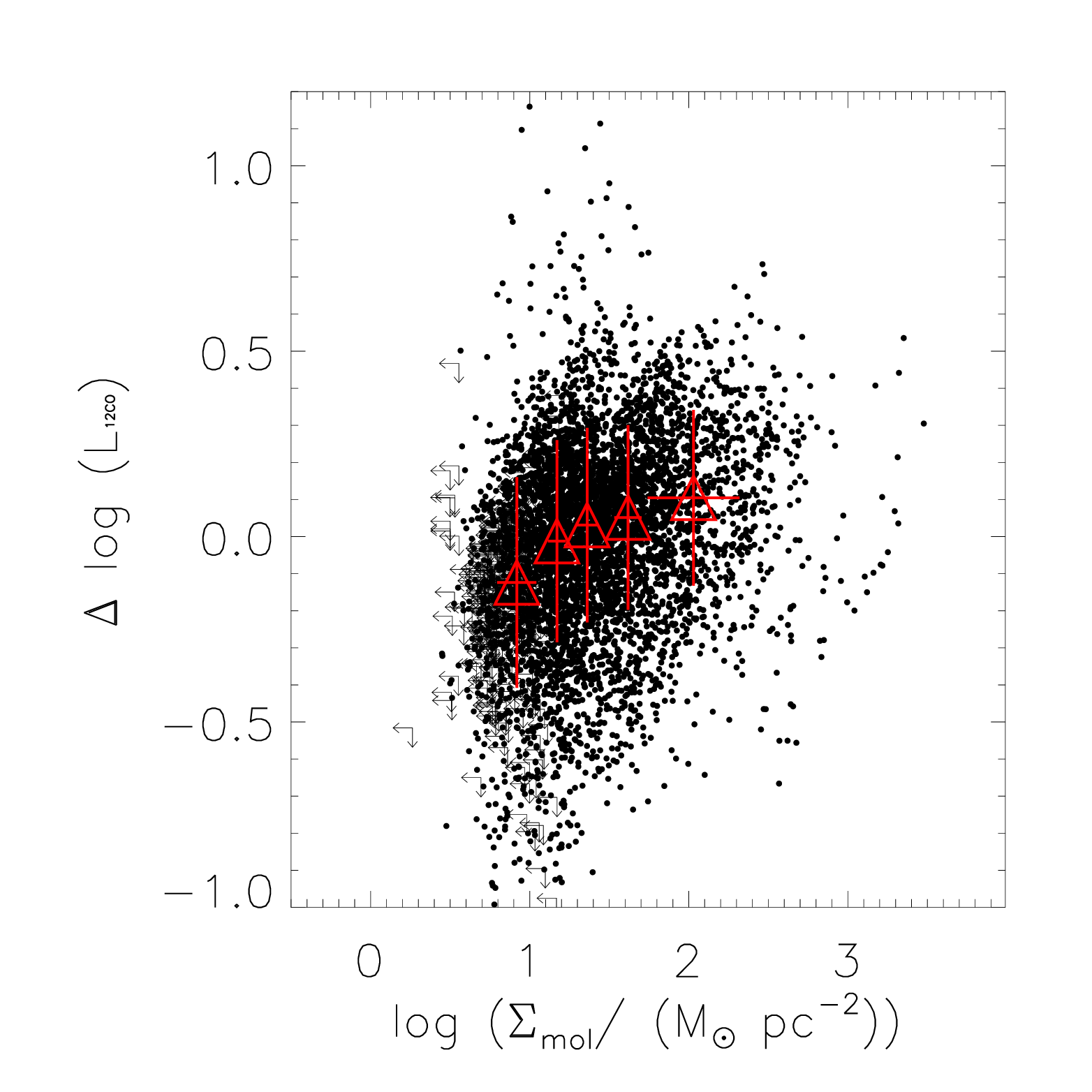}
          		
\caption{The residuals about the \COtw\ versus 12\micron\ luminosity relation (Equation~\ref{eq:resids}) as a function of derived parameters (\COth\ optical depth $\tau(^{13}$CO), gas kinetic temperature T$_{k}$ and molecular gas surface density $\Sigma_{\rm mol}$).} 
\label{fig:off_12}
\end{figure*}
\begin{figure*}
\centering \includegraphics[width=0.32\textwidth]{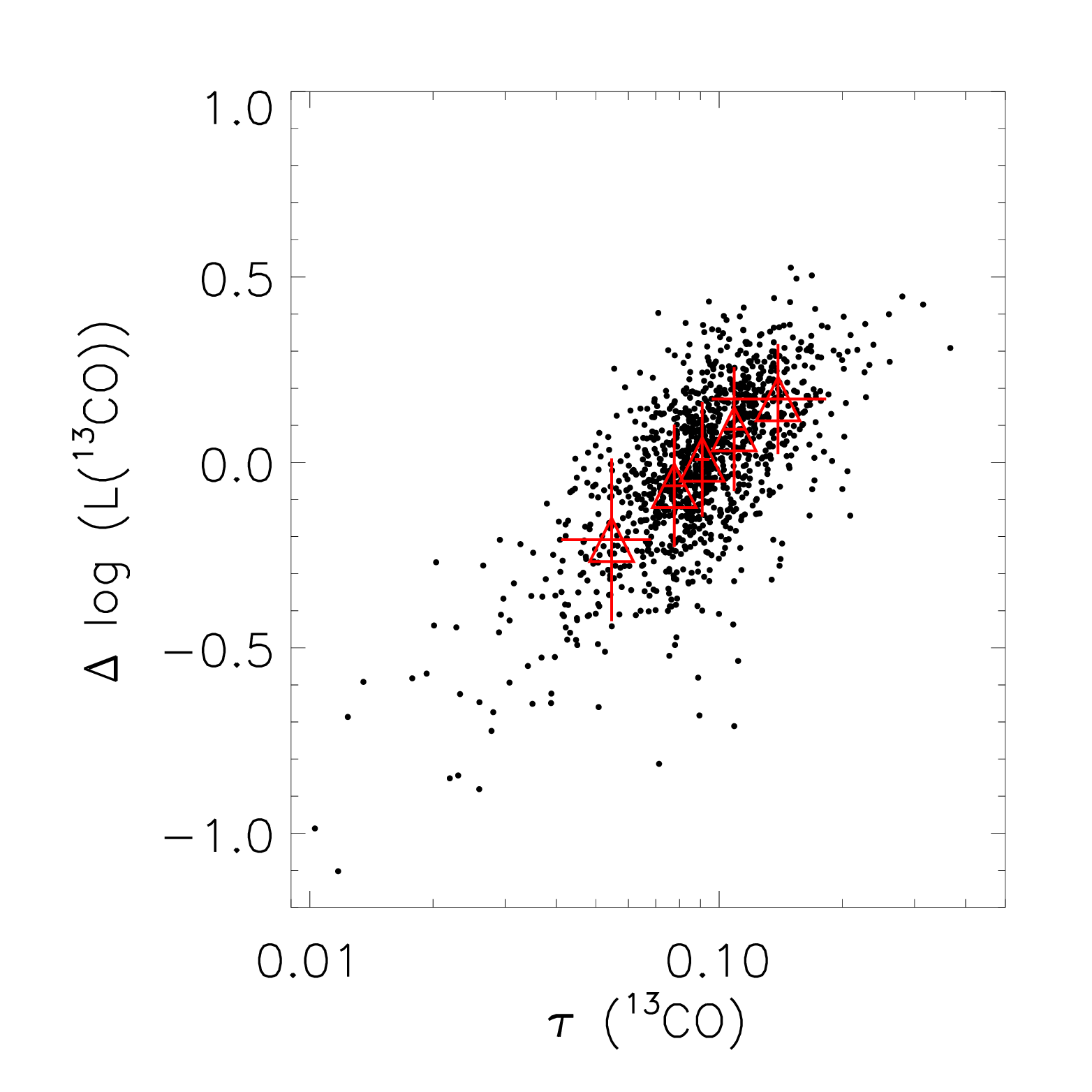}
\includegraphics[width=0.32\textwidth]{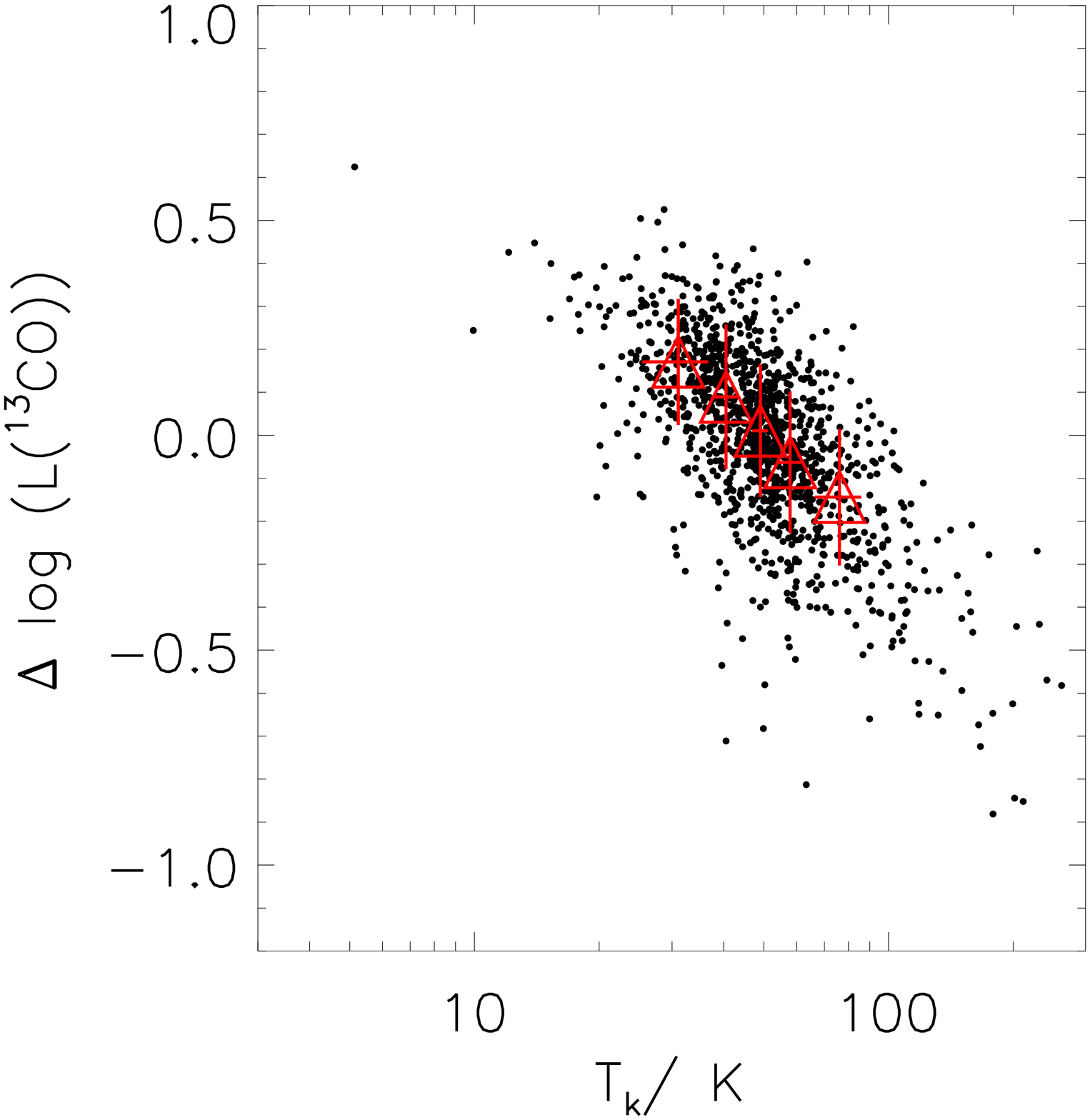}
\includegraphics[width=0.32\textwidth]{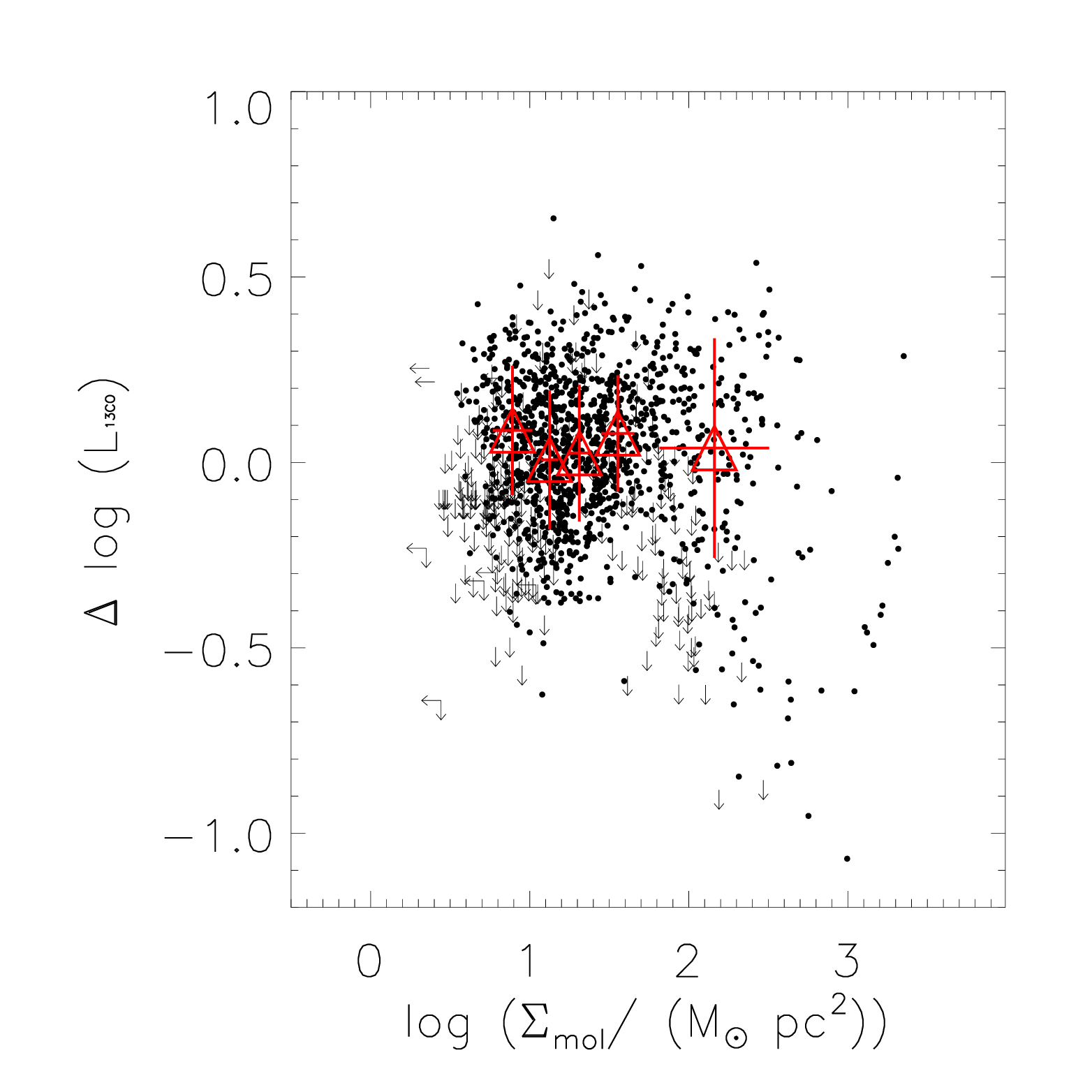}
\caption{Same as Figure~\ref{fig:off_12}, but for the residuals of the predicted value from the best-fitting relation between $L_{^{13}\rm CO}$ and
    $L_{12 \micron}$.} 
\label{fig:off_13}
\end{figure*}

We use the best-fitting relations (Eq.~\ref{co_12} and Eq.~\ref{co_13}) between $L_{^{12}\rm CO}$ or $L_{^{13}\rm CO}$ and  $L_{12 \micron}$ to predict CO luminosity from W3 luminosity, then we compute residuals about these fits, e.g. for the \COtw-12 $\mu$m relation, 
\begin{equation}\label{eq:resids}
    \Delta \log(L_{^{12}\rm CO}) \equiv \log(L_{^{12}\rm CO,obs})- \log(L_{^{12}\rm CO,est}),
\end{equation}
and likewise for \COth\ residuals.
Based on more than one thousand of pixels with both \COtw\ and \COth\ detections, we plot the residuals as a function of \COth\ optical depth and gas kinetic temperature in the left two panels of Figure~\ref{fig:off_12} and Figure~\ref{fig:off_13}. In each panel, to show the trend, we divide the galaxies into four subsamples (with the same number of galaxies in each) according to the parameter considered, and compute the median and scatter for the galaxies in different bins, which is plotted as the large red triangles and error bars. Then we can see the \COtw\ luminosity residuals show nearly no dependence on the optical depth or gas kinetic temperature with the Spearman's correlation coefficient $r$ of -0.1 and 0.12 respectively, while the residuals of $L_{^{13}\rm CO}$ is enhanced when the optical depth increase and gas kinetic temperature decrease ($r \sim$ 0.6 and -0.6). These can lend support to a guess that the relation between \textit{WISE} 12\micron\ band (PAH) and \COtw\ is more physically fundamental. And the relation between 12\micron\ and \COth\ emission, most if not all, is the consequence of relation between 12\micron\ and \COtw\ with the effect of optical depth and temperature.

In Figure~\ref{fig:off_12}, we use the linear fit shown in Figure~\ref{fig:co_w} to predict the \COtw\ luminosity, because it's based on a more complete sample of pixel and more consistent with the global one provided in \citet{Gao2019}, compared with the one in the right panel of Figure~\ref{fig:co2_w}. Then we also check the correlations between  the \COtw\ luminosity residuals and the spatially resolved parameters using the second fitting, and find no correlation (${\lvert}r{\rvert} \leq 0.1$), which support the different dependence between ${\Delta}$ log($L_{^{12}\rm CO})$ and ${\Delta}$ log($L_{^{13}\rm CO})$ on these parameters is real instead of sample selection bias.

We find ${\Delta}$ log($L_{^{12}\rm CO})$ tend to be negative in the region with very low  molecular gas mass surface density shown in the right panel of Figure~\ref{fig:off_12}, which also explain a lot of upper-limits are not above the best-fittng line in  Figure~\ref{fig:co_w}. The Spearman's correlation coefficient ($r$) is 0.34 for all pixels, which is mainly contributed by the regions with $\Sigma_{\rm mol} < 10 {\rm M_{\odot} pc^{-2}}$, and this trend disappears when we just include the pixels with $\Sigma_{\rm mol} > 20 {\rm M_{\odot} pc^{-2}}$ ($r$ = 0.1 for 2623 detected pixels)). 

But, we can't provide the exact range of  $\Sigma_{\rm mol}$ value where the \COtw\ luminosity would be significantly underestimate using 12 \micron\ band, because the CO-to-H$_2$ conversion factor ($\alpha_{\rm CO}$ or $X_{\rm CO}$) can be largely different from the value we assumed in Section~\ref{sec:co}, which is dependent on the physical conditions in the molecular clouds, $X_{\rm CO} \propto n^{0.5}T_{\rm K}^{-1}$, even ignoring metallicity effects and assuming CO emission from the gravitationally bound and virialized cloud cores \citep{Maloney1988,Bolatto2013}.  
Non-correlation between \COtw\ luminosity residual on mass surface density at the mid-to-high end is consistent with independence of \COth\ luminosity residual shown in the right panel of
Figure~\ref{fig:off_13}, because the regions with \COth\ measurements are more likely with high molecular gas mass surface density.

So we guess the correlation between \COtw\ and 12 \micron\ emission should be sustained as long as the molecular gas mass density is not too low, in other words, star formation could take place, where the 12 \micron\ emission is dominated by the PAH formed in molecular clouds.
We also examine the correlation between the offset and the dense gas fraction, which is indicated by the luminosity ratios of HCN(4-3) and HCO$^{+}$(4-3) over \COtw. We find that the offset appears to be nearly independent (with correlation coefficient less than 0.2), because there should be a large amount of molecular gas existing if the dense gas can be detected. 

All these results support that the \COtw\ estimators  based on 12 \micron\ emission is usable in the regions with $\Sigma_{\rm mol}$ higher than a threshold.

\section{Discussion}
\label{sec:discussion}
Compared with other gas tracers, the tight relation between \COtw\ and 12 \micron\ emission is always present, so long as the molecular gas mass surface density is not too low, and would not be affected by the physical conditions of molecular gas such as optical depth and kinetic temperature. We explore the physical origins of this correlation in Sec~\ref{sec:phy_12co_w3}, we test whether the relations are statistically robust from galaxy to galaxy in Sec~\ref{sec:diff_sample}, and we investigate the impact of AGNs on the 12 \micron-\COtw\ 
relation in Sec~\ref{sec:diff_agn}.

\subsection{Why is 12 \micron\ Emission Tightly Linked with \COtw?}
\label{sec:phy_12co_w3}
Our findings show that there is a tight and strong correlation between \COtw\ and 12 \micron\ band emission at scales of about 1 kpc or less. By combining with the global and resolved studies presented by \citet{Jiang2015,Gao2019,Chown2021}, this linear
relation holds over 5 orders of magnitude, and is highly applicable to different physical scales, from spatially resolved regions to whole galaxies. But the physical origin of this relation is still debated.

Our comparisons of the correlations between 12 $\mu$m emission and various molecular species may 
suggests that emission in \textit{WISE} 12 \micron\ band is mainly from the cold dust (mostly PAHs).

On the one hand, in molecular clouds, PAH molecules and \COtw\ are spatially well-mixed, and are excited/destroyed under similar conditions. 
Dense gas is found deeper inside molecular clouds, while PAHs are found primarily on the surfaces of clouds, leading to a weaker association between 12 $\mu$m emission and dense gas.
Considering the \COth\ is mostly optically thin (with median optical depth about 0.09), \COth\ emission originates in denser cores than \COtw\ as suggested in \citet{Mao2000}, which lead to the larger scatter between \COth\ and 12 $\mu$m emission than that between \COtw\ and 12 $\mu$m emission. The slope of ${\Delta}$ log($L_{^{13}\rm CO})$-log(T$_{k}$) is also consistent with the power-law index (-1.4 ) of \COth/\COtw\ intensity ratio as a function of gas kinetic temperature for high densities presented by homogeneous cloud model in \citet{Paglione2001}.
 Similarly we also find the scatter of correlation between \CeiO\ and 12 \micron\ is significant larger than that of \COtw\ over the same set of pixels (37 pixels in 13 galaxies).

 On the other hand, the PAH emission can be detected in wider varieties of astrophysical regions, even in early-type galaxies, where the PAH spectra are dominated by the 11.3 and 12.7 \micron\ band \citep{Kaneda2005,Kaneda2008,Vega2010}. 
  So the PAH emission from the atomic interstellar medium (ISM) not traced by CO can explains why the observed \COtw\ -12 \micron\ correlation breaks down in more diffuse gas, which are also shown as the pixels with very low molecular gas mass surface density in Figure~\ref{fig:off_12}.  This kind of regions or galaxies with few molecular clouds and/or very low molecular gas mass surface density will also be analysed in our next work based on a large sample of nearby early-type galaxy.  
  
  Overall, it is plausible our finding support that the 11.3 \micron\ PAH emission is highly correlated with \COtw\ emission in both position and flux, which is the main reason why \textit{WISE} 12 \micron\ band emission can be used to estimate the molecular gas. Future work could confirm this with a study on smaller physical scales to reveal more detail about PAH formation/destruction and emission mechanisms in  molecular clouds. In the future, we will do some analysis based on some observations in Milky Way \citep[e.g. Milky Way Imaging Scroll Painting;][]{Su2019}, and focus on 11.3 \micron\ PAH that were obtained with the \textit{Spitzer Space Telescope} or with the \textit{James Webb Space Telescope}.

\begin{figure*}
\centering \centering \includegraphics[width=0.24\textwidth]{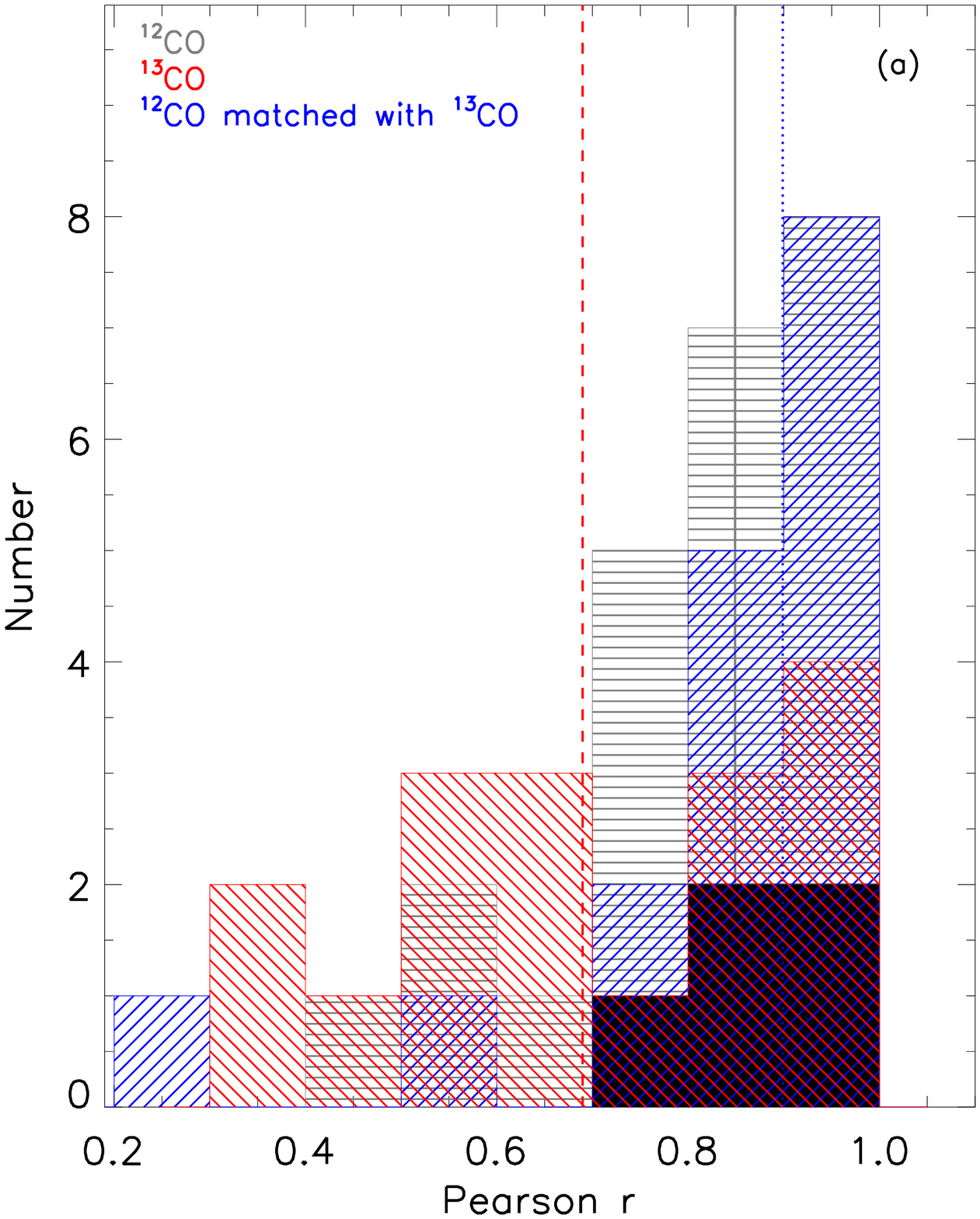}
\centering \centering \includegraphics[width=0.24\textwidth]{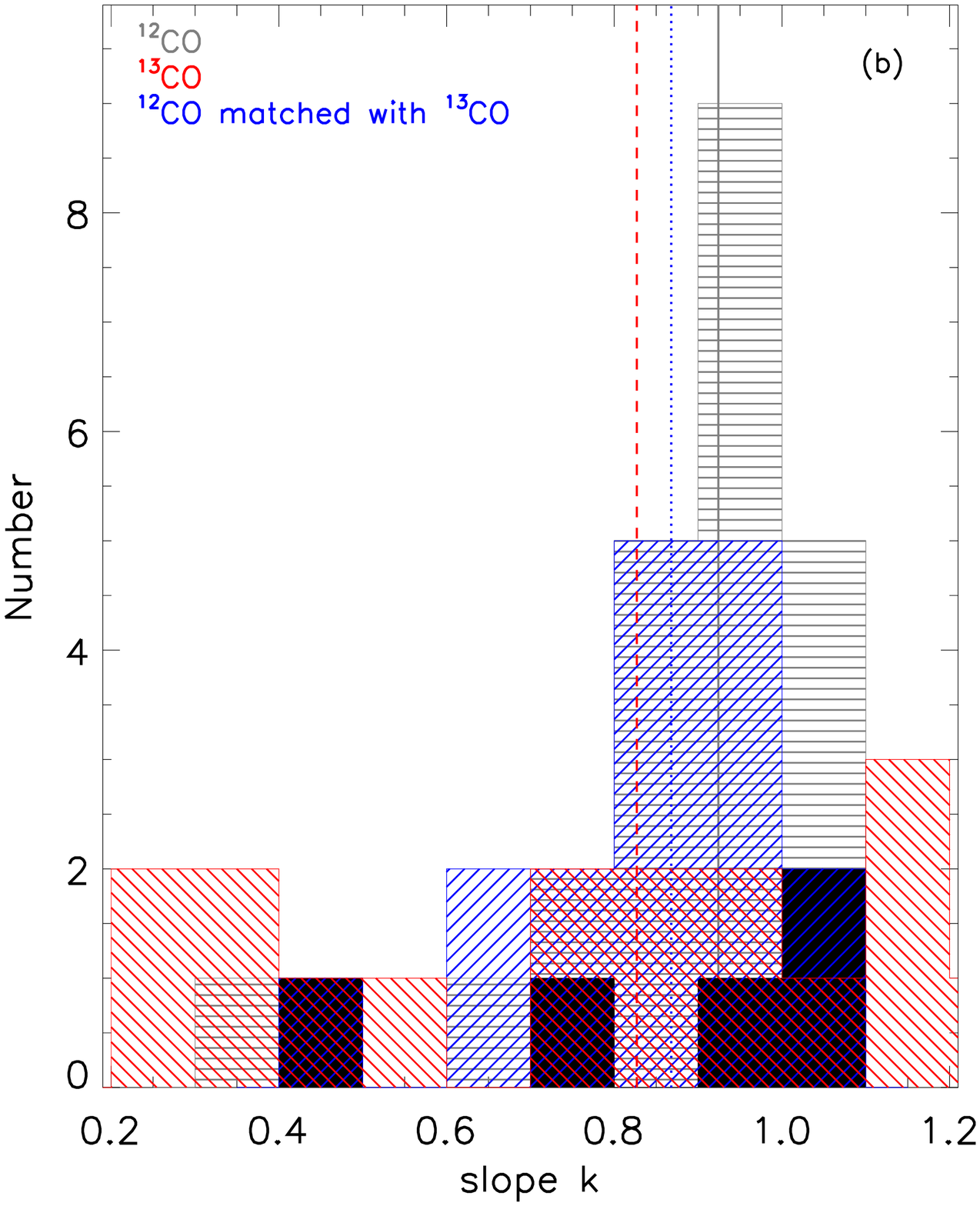}
\centering \centering \includegraphics[width=0.24\textwidth]{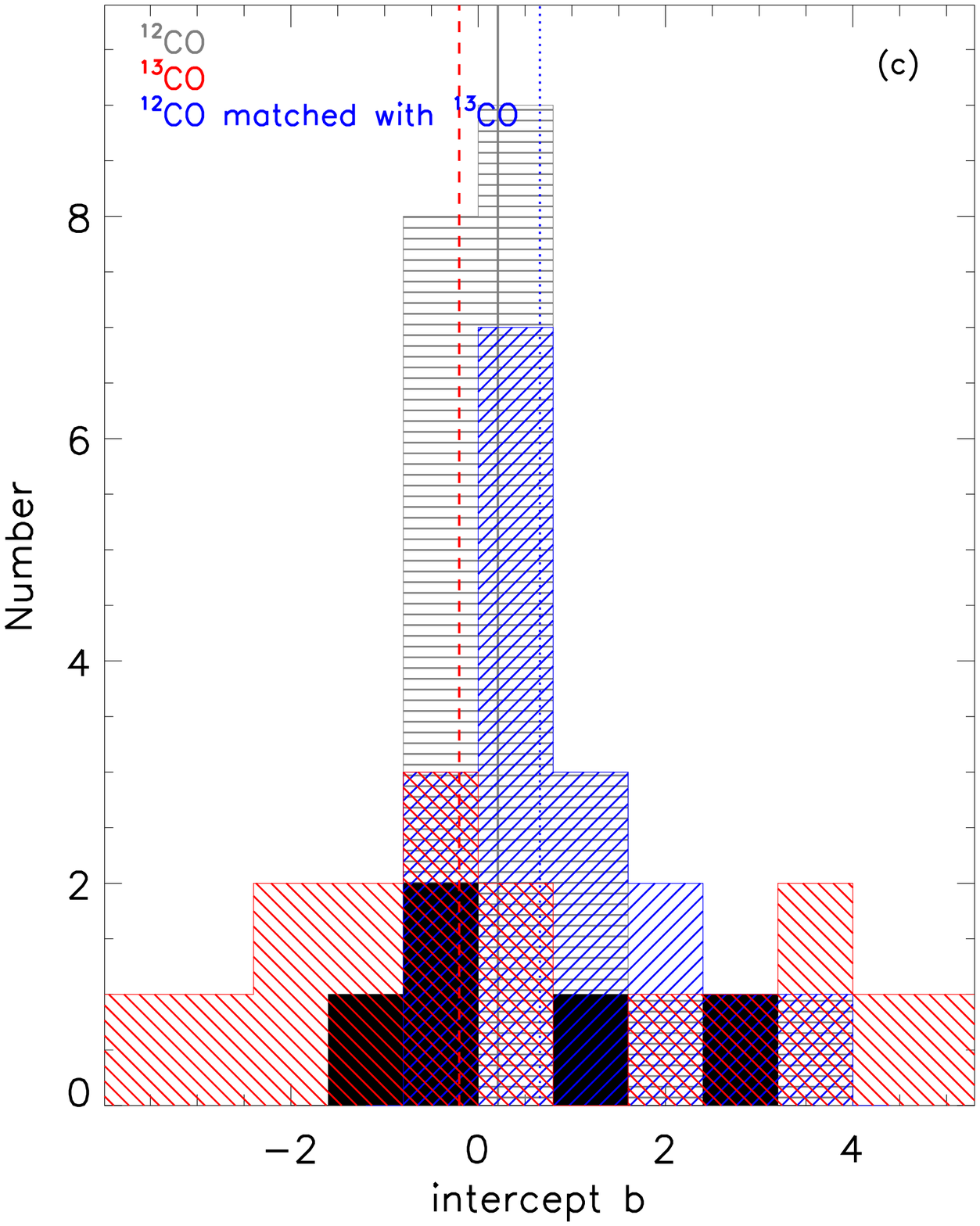}
\centering \centering \includegraphics[width=0.24\textwidth]{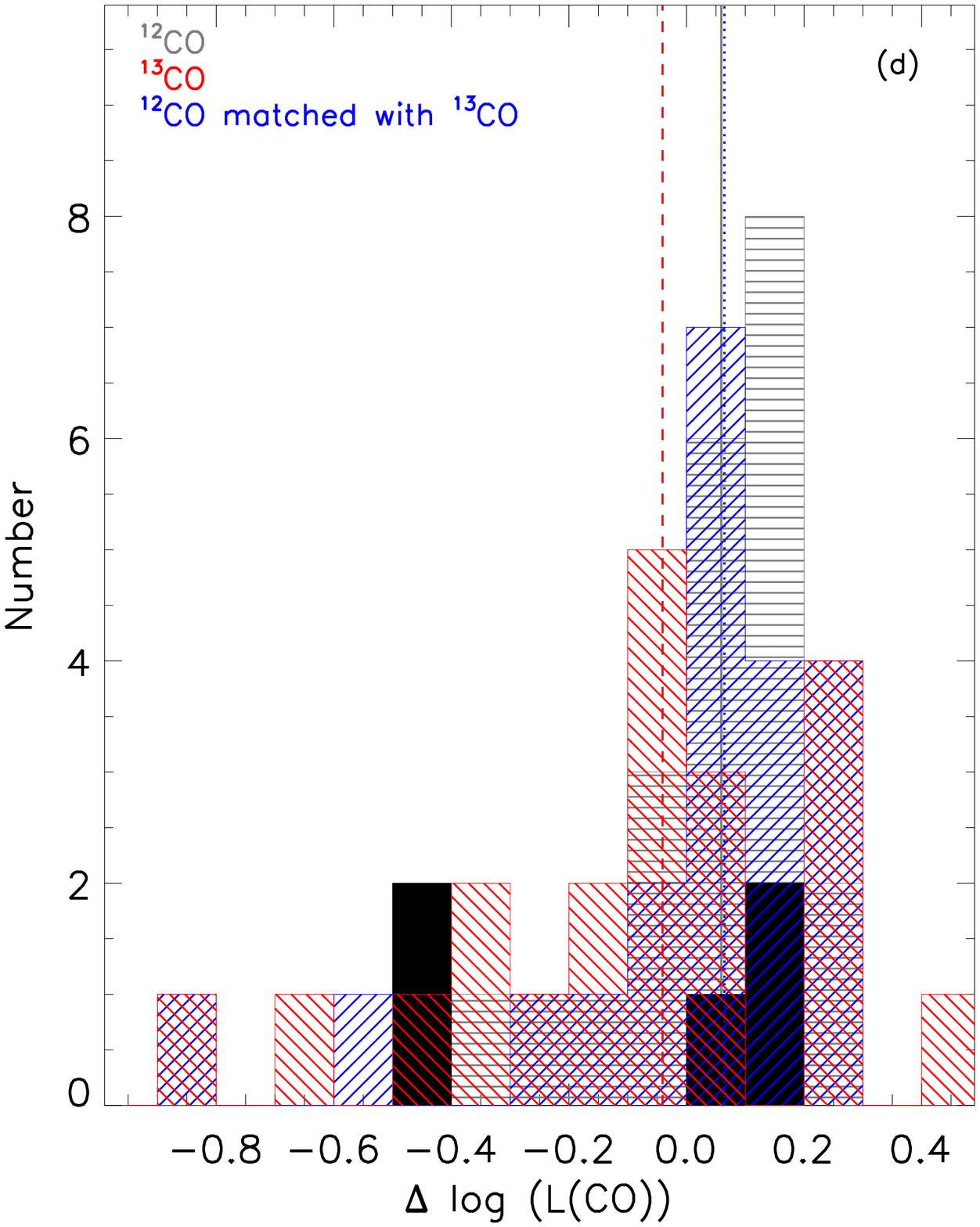}
\caption{The distribution of Spearman's correlation coefficient (a), best-fit slope (b) and intercept (c) between luminosities of 12 \micron\ band (x-axis) and CO isotopologues (y-axis), which is computed based on individual pixel measurements in each galaxy. We only include the galaxies with at least 12 CO detections or valuable upper limits. In the first three panels, the grey histograms show the parameter of log$L_{12 \micron}$ versus log$L_{^{12}\rm CO (1-0)}$ relation obtained for 24 galaxies, and the black ones indicate AGNs, while the ones with red and blue lines show the parameters of log$L_{12 \micron}$ versus log$L_{^{13}\rm CO (1-0)}$ and log$L_{12 \micron}$ versus log$L_{^{12}\rm CO (1-0)}$ based on the measurements of \COth\ detected or tentatively detected pixels in 17 galaxies. In panel (d), we show the distribution of the median ${\Delta}$ log($L_{\rm CO})$ in each galaxies, based on the log($L_{^{12}\rm CO,est}$) from Eq.~\ref{co_12} for 24 galaxies and log($L_{^{13}\rm CO,est}$) from Eq.~\ref{co_13} for 21 galaxies. The coloured vertical lines indicate the medians.   } 
\label{fig:Pearson_and_slop}
\end{figure*}

\subsection{Comparing the 12 \micron-CO Relations in Different Samples}
\label{sec:diff_sample}
To explain the variation in 12 \micron-CO relations, we explore the source of variation and offset from the main relations from galaxy to galaxy, beside the study presented above about the spatially resolved physical conditions.

For that, we first do the similar fitting as Figure~\ref{fig:co2_w} for all galaxies with a minimum 12 measured pixels (detections or valuable upper limits), to calculate the Spearman rank correlation coefficient $r$, slope $k$ and intercept $b$ between log$L_{\rm CO}$ and log$L_{12 \micron}$ for each galaxy.
Figure~\ref{fig:Pearson_and_slop} shows the histograms of these fitting parameters for different isotopologues.

We discuss the effect of sample selection on the best-fit relations (logarithmic) between 12 \micron\ and $\rm CO (1-0)$ luminosity.
In the left panel of Figure~\ref{fig:Pearson_and_slop}, we can find about the 12 \micron\ vs. \COtw\ relation, most galaxies are with a very high Spearman rank correlation coefficient regardless of how to select the pixels used in fitting ( 0.85 $\pm$ 0.15 and 0.90 $\pm$ 0.19 for all \COtw\ measured pixels and \COth\ measured pixels). Then we check the details of 7 galaxies with low value ($r \leq 0.8$), and find the reasons for poor correlations are few CO-detected pixels in such galaxies, or small parameter range in the pixels that are detected (e.g. the $L_{12 \micron}$ region covered by the pixels in most such galaxies is less than 1 dex). 
So it is encouraging to see that the strong correlations between 12 \micron\ and \COtw\ luminosity exist in most galaxies with enough statistical data. 

In contrast, the distribution of Spearman rank correlation coefficient between log$L_{12 \micron}$ and log$L_{^{13}\rm CO (1-0)}$ luminosity are much lower (0.69 $\pm$ 0.2). 
Similarly, the scatter of slope ($k$) and intercept ($c$)  ($\sigma_{k}=0.34$ and $\sigma_{c}=2.6$) is significantly larger than the one 
between 12 \micron\ and \COtw\ ($\sigma_{k}=0.15$ and $\sigma_{c}=1.1$ dex for \COth\ measured pixels) as expected. 

Meanwhile we find: due to the effect of measurement uncertainty, the slope of best-fitting between 12 \micron\ and \COtw\ luminosity tend to be smaller, and the intercept is higher, for galaxies without enough statistical CO pixels, which is similar as the different distribution of $k$ and $c$ fitted using \COtw\ and \COth\ pixel samples. The selection effect combining the AGN contribution may explain the small slope in the right panel of Figure~\ref{fig:dens_w}.   
And we could speculate that the intrinsic 12 \micron-\COtw\ relation (without selection bias) should be nearly linear (with a slope closer to unity).

To assess the applicability of the best-fitting linear relation in different galaxies, we compute the median of the residuals about them for each galaxy.
In the panel (d) of Figure~\ref{fig:Pearson_and_slop}, we show the histograms of median ${\Delta}$ log($L_{\rm CO})$: the histogram with grey lines is computed using Eq.~\ref{co_12} based on all \COtw\ detected pixels in 24 galaxies, and the red and blue one is computed using Eq.~\ref{co_13} and Eq.~\ref{co_12} respectively, based on all \COth\ detected pixels in 21 galaxies. Overall, both estimators can work well for most galaxies, though the scatter is large in the distribution of ${\Delta}$ log($L_{^{13}\rm CO})$.
Based on the grey histogram, we can see the most of median \COtw\ offset is between -0.3 to -0.3, except only 3 galaxies with ${\Delta}$ log($L_{\rm ^{12}CO}/[{\rm K\;km\;s}^{-1}\;{\rm pc^{2}}]) < -0.3 $. Two of these three outliers are AGNs: NGC1068 and NGC4736 (-0.49 dex and -0.42 dex),about  which we will discuss in Sec~\ref{sec:diff_agn}). So the estimator (Eq.~\ref{co_12}) can predict reliable $L_{\rm ^{12}CO}$ for all MALATANG galaxies, even for the incomplete \COth\ pixel sample (the blue histogram).

\subsection{Impact of AGNs on the 12 \micron-\COtw\ Relation}
\label{sec:diff_agn}

The AGN-dominated (Seyfert) galaxies show a larger scatter in global $L_{\rm CO(1-0)}$--$L_{\mbox{12\micron}}$ relation \citep{Gao2019}, where PAHs could survive in the highly ionized medium and show surprising abundant features \citep[e.g.][]{Roche1991,Tommasin2010}.
On the one hand, PAH features at 6.2, 7.7, and 8.6 \micron\ are often substantially suppressed in AGN compared to star-forming galaxies \citep[e.g.][]{Valiante2007,Sajina2008,Diamond-Stanic2010}, which is interpreted as the selective destruction of small PAHs molecules by the hard radiation field of AGNs \citep{Smith2007}. On the other hand, the 11.3 \micron\ PAH features can be  excited by photons or the radiation field from AGNs itself or circumnuclear star formation \citep{Jensen2017}. \citet{Li2020} speculates that the 11.3/7.7 ratios observed in a number of Seyferts could be produced by catacondensed PAHs with an open, irregular structure (be able to have more H atoms on a per C atom basis).
And all these features may appear weaker, due to increased dilution from the AGN continuum \citep{Alonso-Herrero2014} or IR (especially in short wavelengths ) emission from dust heated by the AGN \citep{Shipley2013}.

There are a considerable amount ($\geq$ 50, and about 1000 in total) of measured \COtw\ pixels for 5 out of 8 AGNs in MALATANG sample, which can be used to determine whether and how AGN affect the 12 \micron\ vs. \COtw\ relation. There are only 3 AGNs in \COth\ sample, and no one with the number of detected pixels $\geq$ 30, so we didn't provide the analysis on them. 

In the Figure~\ref{fig:Pearson_and_slop}, we can see that all AGNs also show high correlation coefficient between 12 \micron\ and \COtw\ emission, the best fit slope and intercept of AGNs tend to locate at the two end, and the \COtw\ estimator (Eq.~\ref{co_12}) still work well basically in the AGN sample. It seems these AGNs would not have significant effect on the derived primary $L_{\rm CO(1-0)}-L_{\mathrm{12\mu m}}$ relationship shown in Figure~\ref{fig:co_w}. A linear least-squares fit to the data points excluding the two AGN outliers (NGC1068 and NGC4736) or all five AGNs yields:
\begin{equation} 
\log L_{\rm ^{12}CO} = (0.96 \pm 0.01) \log L_{12 \micron}+(0.00 \pm 0.04),
\end{equation}
 which is closer to the relation of global galaxies \citep{Gao2019}.

In the two AGN hosts with significant \COtw\ residuals, the pixels in these galaxies are well-fit with a linear relation that is approximately parallel to the best-fit relation in Eq.~\ref{co_12}, which covers an extensive radius from the center: 90$''$ ($\sim$6 $r_{K,e}$, corresponding to a physical scale of $\sim$ 4.4 kpc) for NGC1068, 60$''$ (about 1.8 $r_{K,e}$, corresponding to $\sim$ 1.3 kpc) for NGC4736, and beyond those regions the uncertainty becomes larger due to few detected pixels. 
The near-infrared $K_s$ band half-light "effective" radii $r_{K,e}$ is taken from the 2MASS Large Galaxy Atlas \citep{Jarrett2003,https://doi.org/10.26131/irsa2}.  
The larger 12 \micron\ luminosities at fixed CO luminosity in AGN hosts compared to galaxies without AGN may be explained by dust-reprocessed emission peaking between rest frame 15 and 60 \micron\ with respect to star-formation dominated galaxies as shown in \citet{Mullaney2011,Salvestrini2022}, and/or by a prevalence of dust grains that are heated by old stars in weak AGN
(predominantly associated with massive, early-type galaxies), which may be contributed by circumstellar dust from AGB stars \citep{Donoso2012,Villaume2015}. So the different behaviors in 12 \micron-\COtw\ relation may provide new constraints on the physical processes behind how AGN affect the surrounding ISM and subsequent galaxy evolution.

\section{Summary}
\label{sec:summary}

Complementing MALATANG sample with literature CO, we investigate scaling relations between 12 \micron\ luminosity and luminosities of various molecular tracers at sub-kpc scales. Then examined the residuals around the best-fitting relations as a function of the (spatially resolved) physical conditions of molecular gas. We do some statistical tests to show the significance of the galaxy-to-galaxy variation and the impact of hosted AGN.
Our main conclusions can be summarized as follows:

\begin{enumerate}
\item We confirm the existence of a strong nearly linear correlation between $L_{^{12}\rm CO}$ and $L_{12 \micron}$ even on sub-kpc scales, and the relation is precisely consistent with global (galaxy-wide) one \citep{Gao2019} when considering upper limits. 

\item Compared with dense gas traces, 12 \micron\ emission is more strongly correlated with CO (the Spearman's correlation coefficient $r \sim$ 0.92 compared to $r \sim$ 0.77 ). The correlation between \COtw\ and 12 \micron\ luminosity has smaller scatter than the relation with \COth\ (the intrinsic scatter $\sigma_{\rm int} = 0.02$ dex compared to $\sigma_{\rm int} = 0.05$).

\item We estimate the $L_{^{12}\rm CO}$ and $L_{^{13}\rm CO}$ offset relative to the best-fitting relations with $L_{12 \micron}$. The ${\Delta}$ log($L_{^{12}\rm CO})$ of pixel with both \COtw\ and \COth\ detections exhibits no correlations with gas kinetic temperature or optical depth, while ${\Delta}$ log($L_{^{13}\rm CO})$ shows strong dependence with correlation coefficient $r$ about 0.6 and -0.6. 

\item The only significant offset in the relation between 12 \micron\ and \COtw\ emission appears in regions with very low gas surface density (about 10 ${\rm M_{\odot} pc^{-2}}$ or less), above which the relation is not affected by the condition of molecular gas.

\item Most AGNs also show strong correlations between 12 \micron\ and \COtw\ luminosity, and the relation is consistent with the best-fit from the entire MALATANG sample, although it shows larger scatter. The most pixels in two AGNs with relatively large offset ($> 0.3$ dex) show an offset linear relation parallel to the one derived from full sample.

\end{enumerate}

We conclude that a tight and strong correlation between \COtw\ and 12 \micron\ emission is highly applicable to different physical scales, linking sub-kpc local regions to whole galaxies. Although the relation varies from galaxy to galaxy,  \COtw\ and 12 \micron\ luminosity overall follow a tight linear correlation over 5 orders of magnitude. Our best-fit relations may be used to estimate the molecular gas masses for regions within galaxies or for entire galaxies. We advise caution when applying the estimator to regions with very low molecular gas mass surface density (e.g. gas poor early-type galaxies) or in a few AGN-dominated (Seyfert) galaxies, where the molecular gas may be overestimated due to increased fraction of PAH emission from atomic ISM not traced by CO or PAH and IR continuum emission excited or heated by AGNs. A demonstration and correction in much more detail will be shown based on a large sample in the next step.

\section*{Acknowledgements}
We thank the anonymous referee for a thorough and helpful report.
We thank the staff at Qinghai Station of the PMO for continuous help with observations and data reduction. This work is supported by the National Science Foundation of China (grant Nos.12033004, 11861131007, 11803090 and 12003070), and the National Key Basic Research and Development Program of China (grant No. 2017YFA0402704).

This publication made use of data from COMING, CO Multi-line Imaging of Nearby Galaxies, a legacy project of the Nobeyama 45-m radio telescope.
\software{STARLINK \citep{Currie2014}, ORAC-DR pipeline \citep{Jenness2015}, GILDAS/CLASS \citep{Pety2005,GILDAS2013}, SExtractor \citep{Bertin1996}, LinMix \citep{Kelly2007}, IDL Astronomy user's library \citep{Landsman1995}}.

\begin{appendix}

\section{The PMO observation}
\label{sec:data_pmo}

To enlarge the CO (especially \COth\ and \CeiO) pixel sample for MALATANG sample, we conducted half-beam-spaced mapping of NGC4736 over about 2$' \times 1.5'$ area, and observations at 3 position along the major axis of NGC5457 using the PMO 14 m millimeter telescope at Delingha. 

\begin{figure*}[]
\begin{center}
\includegraphics[scale=0.75]{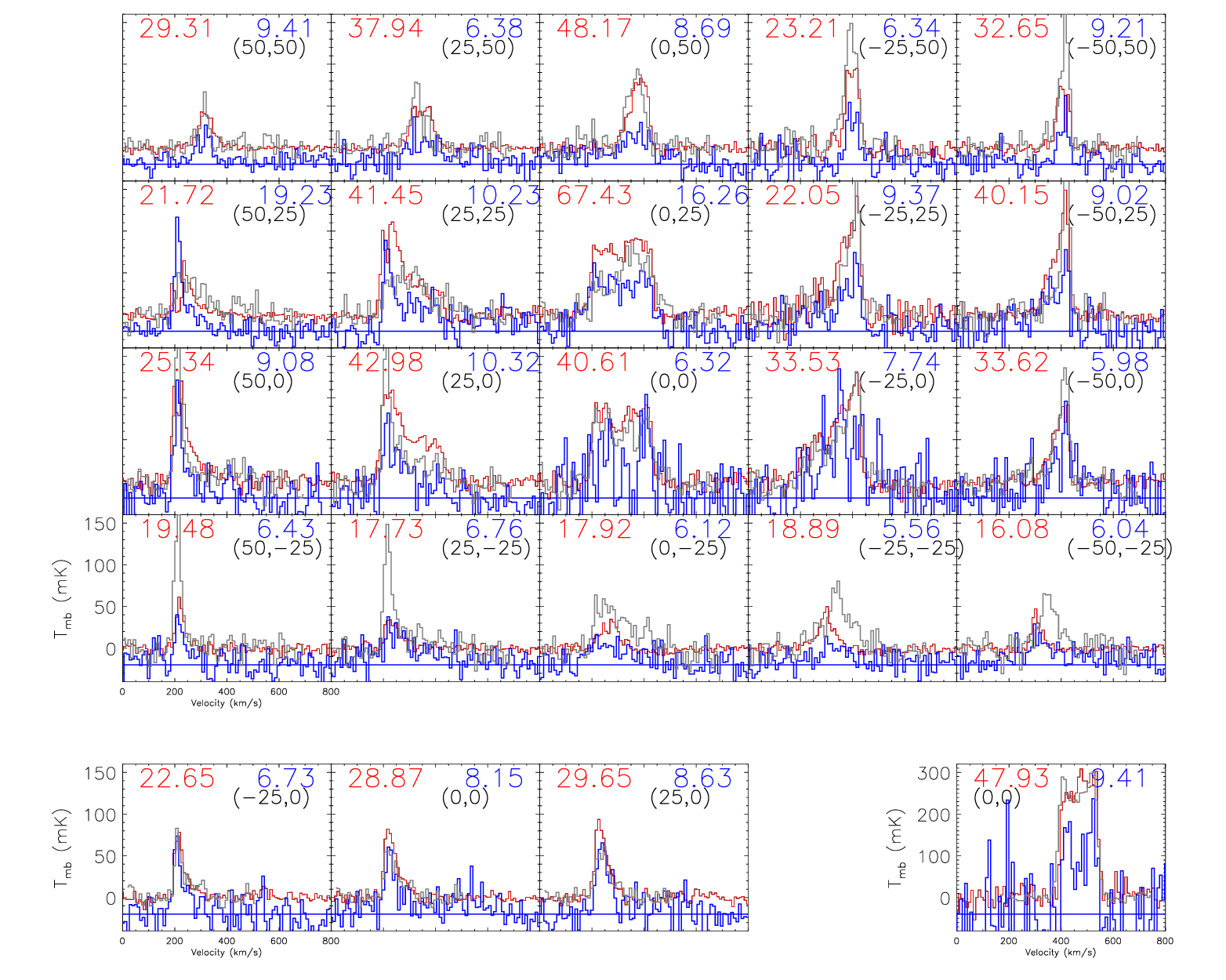}
\caption{ Spectra of \COtw\ (red thin lines) and \COth\ (blue thick lines) in the regions of NGC4736 (top), NGC5457 (lower left) and M51 (lower right) observed with PMO. The temperature of \COth\ is timed by 5 for directly comparison. 
We also derive the simulated PMO \COtw\ spectra (grey thick lines), from the literature one \citep{Kuno2007} based on the parameters of PMO and NRO45M telescope. 
All spectra are on the $T_{\rm mb}$ scale and smoothed to a velocity channel width of 
$\sim$ 10 km s$^{-1}$ for display. The SNR of \COtw\ and \COth\ emission, and offset from the center of belonged galaxy are indicated in each box.}
\label{fig:spe_pmo}
\end{center}
\end{figure*}

To take full advantage of the Galactic dead time, we use the ${\rm ^{12}CO(2-1)}$ maps provided by \citet{Leroy2009} to derive ${\rm ^{13}CO(1-0)}$ integrated intensity maps for galaxies without \COth\ data in the northern sky assuming $R21$ as 0.8 and \COth/\COtw\ as 0.1, and convolve them with the convolution kernel provided by \citet{Aniano2011} to match the beam $\theta_{\rm HPBW} \sim 50''$ of PMO. Then we scale the image by a factor of ${\rm 1.133 \times (HPBW/pixel size)^2}$ and use a conversion factor of 24.9 Jy K$^{-1}$ to compute the estimated velocity-integrated line intensities in $T_{\rm mb}$ scale observed with PMO beam for the positions from the center to the outer disk separated by half-beam size. Finally, we select these 23 positions in the two galaxies with predicted on-source time for \COth\ to reach $S/N = 5$ not exceed 15 hrs by assuming that a line width of emission of 100 km s$^{-1}$. The rms noise level in the spectra with integration time of 1 minute and smoothed to 20 km s$^{-1}$ is 20 mK (computed based on the observation in \citealp{Gao2019}). Besides, we re-observed the center of M51 to compare with the spectrum provided by \citet{Tan2011} to check the effect of upgraded instrument on calibration.

Our observations were carried out with a 3 $\times$ 3 multibeam sideband separation superconducting receiver \citep{Shan2012} from October 2020 to January 2021, and the total on-source time is $\sim$ 128 hours. Two of the nine beams were used simultaneously, one covering the target and one pointing to an off-target area, and the SIS receiver is in double-sideband mode, to observe the emission lines of three CO isotopologues simultaneously, \COtw\ in the upper sideband, and \COth\ and \CeiO\ in the lower sideband.

We use the CLASS package to reduce the data, which is part of the {\tt GILDAS} software package \citep{Pety2005}. After replacing the channels with abnormally strong features by the average flux of the neighboring channels, we visually inspect all scans and classify them into 4 types: good, average, poor and reserved for deleted, based on the quality. First, we just use the "good" scans (with flat baselines, small noise, and no anomalous feature) to obtain an averaged spectrum, and then compute the integrated intensity and uncertainty as introduced in \citet{Gao2019} based on the CO-emitting velocity ranges determined from \COtw\ spectra, because \COtw\ lines are detected at high $S/N$ in all position. 
The worse scans usually would not affect the final integrated intensity tremendously, but maybe helpful to reduce the noise in final spectra, because their rms is large and their weight is low when stacking. So for getting spectra with higher $S/N$, we try to include scans with worse quality after subtract the distorted baseline until the difference of integrated intensity is larger than the uncertainty. Then we select and show the final \COtw\ spectra with highest $S/N$ from 4 tests in Figure~\ref{fig:spe_pmo}. 
And there is one more test of \COth\ spectra is just using the scans observed simultaneously with the selected \COtw\ scans.  

Both \COth\ and \COtw\ emission are detected in all 23+1 positions in 2+1 galaxies with $S/N$ > 5, with integrated intensity, luminosity (computed following Section~\ref{sec:co}) and corresponding uncertainty listed in Table~\ref{tab:data_pmo}.  But we didn't detect \CeiO\ in any position, so we stack the CO spectra over central 2$' \times 1.5'$ area of NGC4736, and get detection in all the three CO isotopologues (total integrated intensity: $I_{^{12}{\rm CO}}$ is 39.9 $\pm$ 0.5, $I_{^{13}{\rm CO}}$ is 6.1 $\pm$ 0.3 and $I_{{\rm C^{18}O}}$ is 1.1 $\pm$ 0.3 K km s$^{-1}$. 

\begin{deluxetable*}{lcllcccccc}
  \tablecaption{Observed and derived properties of \COtw\ and \COth\ for galaxies observed with PMO}
\tablewidth{0pt}
  \label{tab:data_pmo}
  \tablehead{
    \colhead{Source}  & 
    \colhead{Offsets} & 
    \colhead{$I_{^{12}{\rm CO}}$}  & 
    \colhead{$I_{^{13}{\rm CO}}$}  & 
    \colhead{Useful Exp. of \COtw }  &
    \colhead{Useful Exp. of \COth }  &
    \colhead{$V_{\rm mean}$}  &
    \colhead{${\Delta}V$} & 
    \colhead{$\log(L_{\rm ^{12}CO})$} &
    \colhead{$\log(L_{\rm ^{13}CO})$}\\
     \colhead{} & \colhead{(arcsec)} & \colhead{(K km s$^{-1}$)} & \colhead{(K km s$^{-1}$)} & \colhead{(min)} & \colhead{(min)} & \colhead{(km s$^{-1}$)} & \colhead{(km s$^{-1}$)}  & \colhead{(${\rm K\;km\;s}^{-1}\;{\rm pc^{2}}$)} & \colhead{(${\rm K\;km\;s}^{-1}\;{\rm pc^{2}}$)}
  }
  \decimalcolnumbers
  \startdata
NGC4736	&	(0,0)	&	18.6	$\pm$	0.5	&	2	$\pm$	0.3	&	60	&	20	&	315	&	250	&	7.33 	$\pm$	0.01	&	6.47 	$\pm$	0.06	\\
	&	(-25,0)	&	16.3	$\pm$	0.5	&	3	$\pm$	0.4	&	66	&	12	&	311	&	256	&	7.25 	$\pm$	0.01	&	6.65 	$\pm$	0.05	\\
	&	(50,0)	&	6.8	$\pm$	0.3	&	1.1	$\pm$	0.1	&	79	&	129	&	234	&	90	&	6.91 	$\pm$	0.02	&	6.16 	$\pm$	0.05	\\
	&	(-25,50)	&	5.8	$\pm$	0.3	&	0.7	$\pm$	0.1	&	119	&	77	&	397	&	90	&	6.89 	$\pm$	0.02	&	6.00 	$\pm$	0.06	\\
	&	(-25,-25)	&	2.5	$\pm$	0.1	&	0.4	$\pm$	0.1	&	421	&	471	&	321	&	104	&	6.46 	$\pm$	0.02	&	5.82 	$\pm$	0.07	\\
	&	(25,0)	&	15.7	$\pm$	0.4	&	1.6	$\pm$	0.2	&	286	&	62	&	298	&	284	&	7.25 	$\pm$	0.01	&	6.39 	$\pm$	0.04	\\
	&	(50,-25)	&	1.9	$\pm$	0.1	&	0.4	$\pm$	0.1	&	253	&	226	&	216	&	49	&	6.36 	$\pm$	0.02	&	5.81 	$\pm$	0.06	\\
	&	(0,-25)	&	2.5	$\pm$	0.1	&	0.6	$\pm$	0.1	&	324	&	253	&	274	&	143	&	6.43 	$\pm$	0.02	&	5.97 	$\pm$	0.07	\\
	&	(0,25)	&	19.1	$\pm$	0.3	&	2.9	$\pm$	0.2	&	174	&	146	&	315	&	250	&	7.37 	$\pm$	0.01	&	6.55 	$\pm$	0.03	\\
	&	(-50,25)	&	8.6	$\pm$	0.2	&	1.1	$\pm$	0.1	&	96	&	141	&	384	&	113	&	6.98 	$\pm$	0.01	&	6.25 	$\pm$	0.05	\\
	&	(-25,25)	&	10.2	$\pm$	0.5	&	1.1	$\pm$	0.1	&	30	&	70	&	382	&	125	&	7.07 	$\pm$	0.02	&	6.22 	$\pm$	0.04	\\
	&	(25,25)	&	13.2	$\pm$	0.3	&	1.8	$\pm$	0.2	&	200	&	152	&	304	&	257	&	7.20 	$\pm$	0.01	&	6.48 	$\pm$	0.04	\\
	&	(-50,0)	&	6.3	$\pm$	0.2	&	1.1	$\pm$	0.2	&	153	&	52	&	378	&	125	&	6.94 	$\pm$	0.01	&	6.21 	$\pm$	0.07	\\
	&	(25,-25)	&	2.4	$\pm$	0.1	&	0.7	$\pm$	0.1	&	298	&	223	&	253	&	125	&	6.54 	$\pm$	0.02	&	6.09 	$\pm$	0.06	\\
	&	(0,50)	&	6.7	$\pm$	0.1	&	0.8	$\pm$	0.1	&	257	&	87	&	369	&	121	&	6.86 	$\pm$	0.01	&	6.07 	$\pm$	0.05	\\
	&	(-50,50)	&	3.5	$\pm$	0.1	&	0.5	$\pm$	0.1	&	309	&	351	&	406	&	68	&	6.58 	$\pm$	0.01	&	5.92 	$\pm$	0.04	\\
	&	(-50,-25)	&	1.9	$\pm$	0.1	&	0.5	$\pm$	0.1	&	371	&	95	&	321	&	102	&	6.43 	$\pm$	0.03	&	5.88 	$\pm$	0.07	\\
	&	(50,25)	&	2.8	$\pm$	0.1	&	1	$\pm$	0.1	&	386	&	408	&	231	&	95	&	6.52 	$\pm$	0.02	&	6.17 	$\pm$	0.02	\\
	&	(25,50)	&	4.4	$\pm$	0.1	&	0.6	$\pm$	0.1	&	696	&	136	&	350	&	121	&	6.69 	$\pm$	0.01	&	5.94 	$\pm$	0.06	\\
	&	(50,50)	&	2.8	$\pm$	0.1	&	0.5	$\pm$	0.1	&	961	&	506	&	321	&	104	&	6.49 	$\pm$	0.01	&	5.82 	$\pm$	0.04	\\
NGC5457	&	(-25,0)	&	3.7	$\pm$	0.2	&	0.8	$\pm$	0.1	&	163	&	88	&	245	&	106	&	6.78 	$\pm$	0.02	&	6.22 	$\pm$	0.06	\\
	&	(0,0)	&	5	$\pm$	0.2	&	0.9	$\pm$	0.1	&	236	&	230	&	264	&	140	&	6.94 	$\pm$	0.01	&	6.22 	$\pm$	0.05	\\
	&	(25,0)	&	5	$\pm$	0.2	&	1	$\pm$	0.1	&	160	&	53	&	242	&	94	&	6.93 	$\pm$	0.01	&	6.33 	$\pm$	0.05	\\
M51	&	(0,0)	&	40.4	$\pm$	0.8	&	4.3	$\pm$	0.4	&	6	&	5	&	472	&	171	&	8.16 	$\pm$	0.01	&	6.88 	$\pm$	0.05	\\
  \enddata
  \tablecomments{ \\
  (1): Galaxy name. \\
  (2): Offsets from the central position listed in Table~\ref{tab:source}. \\
  (3) and (4): The measured integrated intensities (in main-beam brightness temperature $T_{\rm mb}$ scale) and associated uncertainties of the \COtw\ and \COth\ emission lines, which are computed using Eq.~\ref{rms_co} based on the spectra shown in Figure~\ref{fig:spe_pmo}. \\
  (5) and (6): The on-source observing time of the final \COtw\ and \COth\ spectra. The process that select the scans is explained in the text. \\
  (7) and (8): The velocity and line width are used to show the range of the CO emission line, instead of the Gaussian-fitted values. \\
  (9) and (10): The luminosity of \COtw\ and \COth\ calculated from integrated intensity using the method introduced in Section~\ref{sec:co}.
  }
 \end{deluxetable*}

\end{appendix}

\end{document}